\documentclass[lettersize,journal]{IEEEtran}
\usepackage{amsmath,amssymb,amsfonts}
\usepackage{algorithmic}
\usepackage{algorithm}
\usepackage{array}
\usepackage[caption=false,font=normalsize,labelfont=sf,textfont=sf]{subfig}
\usepackage{textcomp}
\usepackage{stfloats}
\usepackage{url}
\usepackage{verbatim}
\usepackage{graphicx}
\usepackage{cite}
\usepackage{xcolor}
\usepackage{cite}
\usepackage[shortcuts,acronym,nomain]{glossaries}
\usepackage{mathtools}
\usepackage{mathdots}
\usepackage{tikz,pgfplots}
\usetikzlibrary{positioning,shapes.geometric,spy,calc}
\pgfplotsset{compat=1.18}
\usepgfplotslibrary{groupplots}
\usepackage{siunitx}
\usepackage{booktabs}
\usepackage{arydshln}
\usepackage[hidelinks]{hyperref}
\usepackage[noabbrev]{cleveref}
\usepackage{upgreek}

\AtBeginDocument{

}

\crefname{equation}{}{}

\graphicspath{{figures/}}

\newcommand{\ve}{\mathbf}
\newcommand{\ave}[1]{\underline{\mathbf{#1}}}
\newcommand{\m}{\mathbf}
\newcommand{\am}[1]{\underline{\mathbf{#1}}}
\DeclarePairedDelimiterXPP\ExpIntern[1]{\mathbb{E}}{[}{]}{}{#1}
\newcommand{\Exp}{\mathop{}\ExpIntern}
\DeclarePairedDelimiterXPP\VarIntern[1]{\text{Var}}{[}{]}{}{#1}
\newcommand{\Var}{\mathop{}\VarIntern}
\DeclarePairedDelimiterXPP\TraceIntern[1]{\text{tr}}{(}{)}{}{#1}
\newcommand{\tr}{\mathop{}\TraceIntern}
\renewcommand{\Re}{\text{Re}}

\newcommand{\Ndft}{{N_{\text{DFT}}}}
\newcommand{\Nsym}{{N_{\text{sym}}}}
\newcommand{\Lcp}{{L_{\text{CP}}}}
\renewcommand{\j}{\ensuremath{\text{j}}}

\newacronym{ASIC}{ASIC}{application-specific integrated circuit}
\newacronym{FID}{FID}{frequency-independent}
\newacronym{FD}{FD}{frequency-dependent}
\newacronym{FIM}{FIM}{Fisher information matrix}
\newacronym{IRR}{IRR}{image-rejection ratio}
\newacronym{ILR}{ILR}{image-leakage ratio}
\newacronym{RV}{RV}{random variable}
\newacronym{RVC}{RVC}{real valued compensator}
\newacronym{CVC}{CVC}{complex valued compensator}
\newacronym{CMOS}{CMOS}{complementary metal–oxide–semiconductor}
\newacronym{LO}{LO}{local oscillator}
\newacronym{DFT}{DFT}{discrete Fourier transform}
\newacronym{IDFT}{IDFT}{inverse discrete Fourier transform}
\newacronym{FFT}{FFT}{fast-Fourier transform}
\newacronym{IFFT}{IFFT}{inverse fast-Fourier transform}
\newacronym{iid}{i.i.d.}{independent and identically distributed}
\newacronym{CDF}{CDF}{cumulative distribution function}
\newacronym{PDF}{PDF}{probability density function}
\newacronym{PDK}{PDK}{process design kit}
\newacronym{PMF}{PMF}{probability mass function}
\newacronym{PSD}{PSD}{power spectral density}
\newacronym{ADC}{ADC}{analog-to-digital converter}
\newacronym{DC}{DC}{direct current}
\newacronym{DCR}{DCR}{direct-conversion receiver}
\newacronym{AWGN}{AWGN}{additive white Gaussian noise}
\newacronym{WGN}{WGN}{white Gaussian noise}
\newacronym{MSE}{MSE}{mean square error}
\newacronym{CRLB}{CRLB}{Cram\'{e}r-Rao lower bound}
\newacronym{CLT}{CLT}{central limit theorem}
\newacronym{I/Q}{I/Q}{in-phase and quadrature-phase}
\newacronym{I}{I}{in-phase}
\newacronym{Q}{Q}{quadrature-phase}
\newacronym{OFDM}{OFDM}{orthogonal frequency-division multiplexing}
\newacronym{OFDMA}{OFDMA}{orthogonal frequency-division multiple access}
\newacronym{RF}{RF}{radio frequency}
\newacronym{QAM}{QAM}{quadrature amplitude modulation}
\newacronym{QPSK}{QPSK}{quadrature phase-shift keying}
\newacronym{BPSK}{BPSK}{binary phase-shift keying}
\newacronym{CA}{CA}{carrier aggregation}
\newacronym{CFO}{CFO}{carrier-frequency offset}
\newacronym{NR}{NR}{New Radio}
\newacronym{DL}{DL}{downlink}
\newacronym{CP}{CP}{cyclic-prefix}
\newacronym{SNR}{SNR}{signal-to-noise ratio}
\newacronym{SMBE}{SMBE}{simplified MBE}
\newacronym{ISMBE}{ISMBE}{iterative SMBE}
\newacronym{AFSMBE}{AFSMBE}{$\alpha$-filtered simplified moment-based estimator}
\newacronym{MBE}{MBE}{moment-based estimator}
\newacronym{CBE}{CBE}{circularity-based estimator}
\newacronym{EWMA}{EWMA}{exponentially weighted moving average}
\newacronym{FPGA}{FPGA}{field-programmable gate array}
\newacronym{3GPP}{3GPP}{3rd Generation Partnership Project}
\newacronym{LUT}{LUT}{lookup table}
\newacronym{UE}{UE}{user equipment}
\newacronym{ULC}{ULC}{ultra-low complex}
\newacronym{GMLE}{GMLE}{Gaussian maximum-likelihood estimator}
\newacronym{GE}{GE}{gate equivalent}
\newacronym{LMS}{LMS}{least mean squares}
\newacronym{NLMS}{NLMS}{normalized least mean squares}
\newacronym{RLS}{RLS}{recursive least squares}
\newacronym{FSS}{FSS}{joint first and second order statistics}
\newacronym{JFSCS}{JFSCS}{joint first order statistics and conjugate signal model}
\newacronym{CORDIC}{CORDIC}{Coordinate Rotation Digital Computer}
\newacronym{ALM}{ALM}{adaptive logic module}
\newacronym{HW}{HW}{hardware}
\newacronym{RB}{RB}{resource block}
\newacronym{BER}{BER}{bit error ratio}
\newacronym{EVM}{EVM}{error vector magnitude}
\newacronym{DSP}{DSP}{digital signal processor}
\newacronym{SGD}{SGD}{stochastic gradient descent}
\newacronym{SVRG}{SVRG}{stochastic variance reduced gradient}
\newacronym{IIR}{IIR}{infinite impulse response}
\newacronym{BFGS}{BFGS}{Broyden-Fletcher-Goldfarb-Shanno}
\newacronym{L-BFGS}{L-BFGS}{limited-memory BFGS}

\makeglossaries

\begin{document}

\title{On the CRLB for Blind Receiver I/Q Imbalance Estimation in OFDM Systems: \\
Efficient Computation and Closed-Form Bounds}

\author{
    \IEEEauthorblockN{
    Moritz Tockner\IEEEauthorrefmark{1}$^1$,
    Oliver Lang\IEEEauthorrefmark{1}$^2$,
    Andreas Meingassner-Lang\IEEEauthorrefmark{1}$^3$,
    Mario Huemer\IEEEauthorrefmark{1}$^4$}\\
\IEEEauthorblockA{\IEEEauthorrefmark{1}Institute of Signal Processing, Johannes Kepler University Linz,     Austria\\
\{$^1$moritz.tockner, $^2$oliver.lang, $^3$andreas.meingassner-lang, $^4$mario.huemer\}@jku.at}
}



\maketitle

\begin{abstract}
  Modern mobile communication receivers are often implemented with a direct-conversion architecture, which features a number of advantages over competing designs. A notable limitation of direct-conversion architectures, however, is their sensitivity to amplitude and phase mismatches between the in-phase and quadrature signal paths. Such \ac{I/Q} imbalances introduce undesired image components in the baseband signal, degrading link performance---most notably by increasing the bit-error ratio. Considerable research effort has therefore been devoted to digital techniques for estimating and mitigating these impairments. Existing approaches generally fall into two categories: data-aided methods that exploit known pilots, preambles, or training sequences, and blind techniques that operate without such prior information. For data-aided estimation, \acp{CRLB} have been established in the literature. In contrast, the derivation of a \ac{CRLB} for the blind \ac{I/Q}-imbalance estimation case is considerably more challenging, since the received data is random and typically non-Gaussian in the frequency domain. This work extends our earlier conference contribution, which introduced a \ac{CRLB} derivation for the blind estimation of \ac{FID} receiver \ac{I/Q} imbalance using \ac{CLT} arguments. The extensions include a computationally efficient method for calculating the bound, reducing complexity from cubic in the number of samples to linear in the \ac{FFT} size, along with a simplified closed-form approximation. This approximation provides new insights into the allocation-dependent performances of existing estimation methods, motivating a pre-estimation filtering modification that drastically improves their estimation performance in certain scenarios.
\end{abstract}


\section{Introduction}
The direct-conversion or homodyne receiver lends itself to popularity in state-of-the-art mobile communication transceiver designs due to multiple advantages compared to a classical heterodyne architecture \cite{razavi_rf-microelectronics_2013}. It operates by splitting the received \ac{RF} signal into two paths: the \acl{I/Q} path. Each path is subsequently downconverted to the baseband using the \ac{LO} signal. However, due to e.g., manufacturing tolerances or component aging, the \ac{I/Q} paths often exhibit unequal signal gains and a phase relation that deviates from the ideal $90^\circ$. This results in gain and phase mismatches (commonly referred to as \ac{I/Q} imbalance) in the resulting complex baseband signal.

These mismatches are typically assumed to be \ac{FID} and lead to an unwanted image in the downconverted signal spectrum. Typically, analog filtering is implemented between the downmixing and the \ac{ADC}, which also results in slight differences between the \ac{I/Q} paths. These lead to \ac{FD} imbalances, which are typically smaller in magnitude than the \ac{FID} imbalances~\cite{schenk_estimation_2006,gottumukkala_optimal_2009,lang_ofdm-based_2024}, and are not considered in this work.

We derive a \ac{CRLB} for the blind estimation problem of an \ac{FID} receiver \ac{I/Q} imbalance based on a general \ac{OFDM} \ac{DL} signal model. The derivation applies to any \ac{CP}-\ac{OFDM} system, encompassing current standards such as 5G \ac{NR} as well as future systems like 6G. The \ac{CRLB} provides a useful reference for unbiased estimators, as it represents the minimum achievable estimation variance of the unknown parameters \cite{kay_fundamentals_1993}.

For \ac{FID} receiver \ac{I/Q} imbalance estimation, \acp{CRLB} are extensively investigated in literature. However, in most studies either a training sequence or pilot subcarriers are required \cite{sakai_joint_2016,namgoong_crlb-achieving_2012,gappmair_crlb_2012,cao_parametric_2006,zou_joint_2006,gil_joint_2005,giugno_efficient_2005}, which entails exact knowledge of specific received symbols at the receiver. It also necessitates the estimation of other non-idealities distorting these known symbols, such as \ac{CFO}, phase-offset or a fading channel. In this work, we consider a fully blind estimation approach without any knowledge of the receive signal, and derive the corresponding \ac{CRLB}.

Related literature includes \cite{meng_joint_2018,liu_joint_2011,paireder_enhanced_2021}. In \cite{meng_joint_2018}, the \ac{CRLB} is derived under a frequency-domain constant-modulus assumption on the transmitted symbols, used as prior knowledge, which restricts the applicability to \ac{QPSK} and \ac{BPSK}. This assumption is fragile in practice, as frequency-selective channels impose frequency-dependent attenuation \cite{proakis_communication_2002}, thereby destroying the constant modulus property in the frequency domain. The formulation further requires joint estimation of \ac{I/Q} imbalance and \ac{CFO}.
Moreover, \cite{meng_joint_2018} reports the \ac{CRLB} as plotted curves and omits both, the derivation, and the closed-form expression. This precludes a reproducible analytical comparison. Furthermore, since the work relies on the constant-modulus assumption, which is misaligned with our general \ac{QAM}-based setting, and no closed-form expression is available for comparison, we exclude the \ac{CRLB} of \cite{meng_joint_2018} from our comparison.

\cite{liu_joint_2011} simplifies the \ac{CRLB} derivation by including the transmitted symbols as additional parameters, making the bound data dependent. A data-independent bound can then be obtained through an ensemble average from multiple simulation runs with random data, significantly increasing the simulation time. Additionally, the computational complexity per run scales cubically with the number of samples used for estimation. In contrast, our method requires no repeated simulations and can be optimized to a computational complexity independent of the number of processed \ac{OFDM} symbols. Hence, it scales linearly with the \ac{FFT} size. 
% Moreover, it yields insights enabling drastically improved estimation performance in specific scenarios. 

In~\cite{paireder_enhanced_2021}, it is shown that for asymmetric subcarrier allocations, the estimation bias of a specific, well-known \ac{MBE}~\cite{bowman_estimation_2006,anttila_blind_2006} is limited solely by the \ac{WGN} added after the imbalance, and by the imbalance itself. In contrast, we provide a similar but more general conclusion, showing via the \ac{CRLB} that this result extends to the \ac{MSE} of any unbiased estimator.

This paper is a substantially extended version of our foundational work presented at the 2025 Asilomar Conference on Signals, Systems, and Computers \cite{tockner_crlb_2025}. While the conference paper introduced the core \ac{CRLB} derivation, this manuscript provides several significant new contributions: (i) a computationally efficient method for calculating the bound, reducing the computational complexity from cubic in the number of samples used for estimation, to linear in the \ac{FFT} size; (ii) a simplified, closed-form approximation of the \ac{CRLB} that yields deeper analytical insights into its dependency on the subcarrier allocation (in particular its symmetry), the number of \ac{OFDM} symbols, and the \ac{SNR}; (iii) a novel pre-estimation filtering technique to improve estimator performance in scenarios where the allocation consists of asymmetric and symmetric components; and (iv) a comprehensive performance analysis across various \ac{SNR} and \ac{ILR} values.

The remainder of this paper is organized as follows: Section~\ref{sec:signal_model} provides an overview of the typically used \ac{FID} \ac{I/Q} imbalance model. Section~\ref{sec:CRLB} presents the statistical signal model and its underlying assumptions, along with a justification for their validity, followed by the derivation of the \ac{CRLB}, its optimized evaluation, and a simplified closed-form approximation. In Section~\ref{sec:simulations}, the estimation performances of multiple state-of-the-art algorithms are evaluated, and their behavior for different simulation scenarios is compared to the \ac{CRLB}. This section also demonstrates the effectiveness of the proposed pre-estimation filtering technique. Finally, Section~\ref{sec:conclusion} concludes the paper.
\section{Signal Model} \label{sec:signal_model}
\begin{figure}[!t]
  \centering
  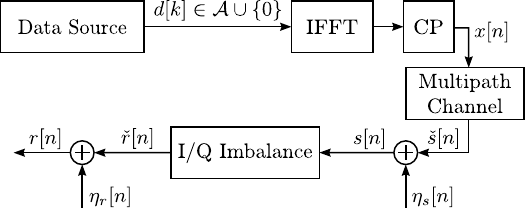
  \caption{Basic block diagram of the equivalent baseband signal processing chain with \ac{I/Q} imbalance.}
  \label{fig:iq_imbalance_bock_diagram}
\end{figure}

In this paper, we adopt a baseband signal model analogous to the one presented in~\cite{tockner_optimum_2024}. An overview of the key processing steps is provided in Fig.~\ref{fig:iq_imbalance_bock_diagram}.
\subsection{Baseband Model}
In this work, we consider a general \ac{CP}-\ac{OFDM} \ac{DL} signal model. A data source generates uniformly distributed random modulation symbols $d[k]$ from a symbol alphabet $\mathcal{A}$ like \ac{QPSK} or higher-order \ac{QAM}. The derivations begin with a single \ac{OFDM} symbol and extend naturally to multiple symbols owing to statistical independence across consecutive symbols.

The $n$th frequency-domain \ac{OFDM} symbol with $\Ndft$ subcarriers is defined as 
\begin{equation} \label{eq:d_n}
  \ve{d}_n = [d_n[0], d_n[1], \dots, d_n[\Ndft - 1]]^T,
\end{equation}
with $d_n[k] \in \mathcal{A} \cup \{ 0 \} \text{ for } k=0, \dots, \Ndft-1$, and $(\cdot)^T$ as the transpose operator. If the $k$th subcarrier of the $n$th \ac{OFDM} symbol is not allocated, it follows that $d_n[k] = 0$. From that, we define a vector $\boldsymbol{\psi}_n \in \{0, 1\}^{\Ndft}$ for the logical allocation pattern of the $n$th \ac{OFDM} symbol with elements
\begin{equation} \label{eq:logical_alloc_pattern}
  [\boldsymbol{\psi}_n]_k = \begin{cases}
    0 & \text{for } d_n[k] = 0 \\
    1 & \text{else,}
  \end{cases}
\end{equation}
where $[\cdot]_k$ refers to the $k$th element of a vector. The total number of allocated subcarriers is $L_s = \boldsymbol{\psi}_n^T \m{1}_{\Ndft}$, where $\m{1}_{\Ndft}$ represents the all-one column vector of length $\Ndft$. For simplicity, we assume a constant subcarrier allocation $\boldsymbol{\psi}_n = \boldsymbol{\psi}$ across consecutively received \ac{OFDM} symbols\footnote{The presented method can however easily be extended to varying subcarrier allocations, at the cost of an increased computational complexity.}. The time-domain \ac{OFDM} symbol calculation with
\begin{equation} \label{eq:tilde_x}
  \tilde{\ve{x}}_n = \m{F}^{-1}_{\Ndft} \ve{d}_n
\end{equation}
uses the unitary $\Ndft \times \Ndft$ \ac{DFT} matrix $\m{F}_{\Ndft}$ with elements
\begin{equation}
  [\m{F}_{\Ndft}]_{k,l} = \dfrac{1}{\sqrt{\Ndft}} \exp(-\j 2\pi k l / \Ndft).
\end{equation}
The imaginary unit is denoted by $\j$. A \ac{CP} of $\Lcp$ samples is prepended to the time-domain vector to avoid possible inter-symbol interference due to channel dispersion. The length $\Lcp + \Ndft$ time-domain \ac{OFDM} symbol with a \ac{CP} is then defined as 
\begin{equation} \label{eq:x}
  \ve{x}_n = \m{D}_{\text{CP}} \tilde{\ve{x}}_n,
\end{equation}
with
\begin{equation} \label{eq:D_cp}
  \m{D}_{\text{CP}} = \left[\begin{array}{cc}
    \m{0}_{\Lcp \times (\Ndft - \Lcp)} & \m{I}_{\Lcp} \\
    \noalign{\vskip 2pt}
    \hdashline
    \noalign{\vskip 2pt}
    \multicolumn{2}{c}{
      \m{I}_{\Ndft}
    }
  \end{array}\right]
\end{equation}
as the $\Nsym \times \Ndft$ \ac{CP} matrix, with $\Nsym = \Lcp + \Ndft$. In this paper, $\m{0}_{N \times M}$ represents the $N \times M$ all-zero matrix, whereas $\ve{0}_{N}$ denotes the all-zero column vector of size $N$, and $\m{I}_{N}$ denotes the $N \times N$ identity matrix. The full signal vector $\ve{x} = [x[0], x[1], ..., x[N - 1]]^T$, with the total number of samples $N = N_{\text{OFDM}} \cdot \Nsym$, can be defined as 
\begin{equation}
  \ve{x} = [\ve{x}_0^T, \ve{x}_1^T, ..., \ve{x}^T_{N_{\text{OFDM}} - 1}]^T,
\end{equation}
with $N_{\text{OFDM}}$ as the number of transmitted \ac{OFDM} symbols.
Similarly, throughout the remainder of this work, all other signal vectors without subscripts also represent a concatenation of consecutive \ac{OFDM} symbols, unless explicitly stated otherwise.

The signal $x[n]$ is transmitted over a frequency-selective fading channel with the impulse response $\ve{h} = [h[0], h[1], ... h[Q - 1]]^T$, where $Q < \Lcp$, yielding the channel-distorted signal $\check{s}[n]$. This convolution is represented by the linear transformation  
\begin{equation} \label{eq:check_s}
  \check{\ve{s}} = \m{H} \ve{x},  
\end{equation}
via the $N \times N$ channel impulse response matrix 
\begin{equation}
  \m{H} = \begin{bmatrix}
    h[0] & 0 & 0 & \hdots & 0 \\
    h[1] & h[0] & 0 & \hdots & 0 \\
    h[2] & h[1] & h[0] & \hdots & 0 \\
    \vdots & \ddots & \ddots & \ddots & \vdots \\
    0 & \hdots & h[2] & h[1] & h[0]
  \end{bmatrix}.
\end{equation}
Adding complex-valued circular \ac{WGN} $\boldsymbol{\upeta}_s \sim \mathcal{N}(\ve{0}_{N}, \sigma^2_{\eta_s} \m{I}_{N})$ from the channel and the pre-\ac{I/Q} imbalance analog components yields the distorted, noisy received signal 
\begin{equation} \label{eq:s}
  \ve{s} = \check{\ve{s}} + \boldsymbol{\upeta}_s. 
\end{equation}  
The \ac{FID} receiver \ac{I/Q} imbalance can be modeled by two complex-valued coefficients \cite{tockner_optimum_2024}
\begin{align}
  K_1 & = \cos (\phi/2) - \j \epsilon \sin (\phi / 2), \\
  K_2 & = \epsilon \cos (\phi / 2) + \j \sin (\phi / 2),
\end{align}
with $\epsilon$ and $\phi$ as the gain and phase imbalance, respectively. Subsequently, the \ac{I/Q} imbalanced signal vector is given by
\begin{equation} \label{eq:check_r}
  \check{\ve{r}} = K_1 \ve{s} + K_2 \ve{s}^*,
\end{equation}
where $(\cdot)^*$ represents the complex conjugation.
Introducing the complex augmented imbalance matrix 
\begin{equation} \label{eq:K}
  \am{K} = \begin{bmatrix}
    K_1 & K_2 \\
    K_2^* & K_1^*
  \end{bmatrix} \otimes \m{I}_N,
\end{equation}
where '$\otimes$' denotes the Kronecker product, the complex augmented \ac{I/Q} imbalanced signal vector $\check{\ave{r}} = [\check{\ve{r}}^T, \check{\ve{r}}^H]^T$, with $(\cdot)^H$ as the conjugate transpose operator, is expressed by the widely linear transformation \cite{adali_complex-valued_2011,lang_knowledge-aided_2018}
\begin{equation} \label{eq:check_aug_r}
  \check{\ave{r}} = \am{K} \, \ave{s},
\end{equation}
with $\ave{s} = [\ve{s}^T, \ve{s}^H]^T$.
The addition of a complex-valued circular \ac{WGN} signal vector $\boldsymbol{\upeta}_r \sim \mathcal{N}(\ve{0}_N, \sigma^2_{\eta_r} \m{I}_{N})$, which represents thermal noise from post-\ac{I/Q} imbalance analog components, leads to the noisy imbalanced signal vector
\begin{equation} \label{eq:r}
  \ve{r} = \check{\ve{r}} + \boldsymbol{\upeta}_r,
\end{equation}
which is subsequently used for estimating the \ac{I/Q} imbalance. The literature often simplifies the \ac{I/Q} imbalance estimation problem to estimating
\begin{equation} \label{eq:alpha}
  \alpha = K_2 / K_1^*,
\end{equation}
and relies solely on $\alpha$ for compensation, as any remaining complex-valued scaling of the signal is handled by the subsequent channel equalizer \cite{anttila_blind_2006}. We adopt this convention throughout this work.

An important performance metric for \ac{I/Q} imbalance is the so-called \ac{ILR} in dB \cite{windisch_blind_2007} 
\begin{equation}
  \text{ILR}_{\text{dB}} = \qty{10}{dB} \cdot \log \left( \dfrac{|K_2|^2}{|K_1|^2} \right) = \qty{10}{dB} \cdot \log \left( |\alpha|^2 \right).
\end{equation}
It represents the power ratio between the unwanted image component $K_2 \ve{s}^*$ and the desired signal component $K_1 \ve{s}$.

\subsection{Residual Image Leakage Ratio}
Given an estimate $\hat{\alpha}$ of $\alpha$, the image component can be suppressed, as described in \cite{paireder_ultra-low_2019}, by calculating
\begin{equation}
  \hat{\ve{s}} = \ve{r} - \hat{\alpha} \ve{r}^*.
\end{equation}
Considering only the signal components from \eqref{eq:check_r}, this yields
\begin{equation}
  \hat{\ve{s}} = \ve{s}(K_1 - \hat{\alpha} K_2^*) + \ve{s}^*(K_2 - \hat{\alpha} K_1^*),
\end{equation}
leading to 
\begin{equation}
  \text{ILR}_{\text{c,dB}} = \qty{10}{dB} \cdot \log_{10} \left( \dfrac{|K_2 - \hat{\alpha} K_1^*|^2}{|K_1 - \hat{\alpha} K_2^*|^2} \right).
\end{equation}
By substituting $K_2 = \alpha K_1^*$ from \eqref{eq:alpha}, the residual \ac{ILR} after compensation becomes
\begin{equation} \label{eq:ILR_c_dB}
  \begin{aligned}
      \text{ILR}_{\text{c,dB}} & = \qty{10}{dB} \cdot \log_{10} \left( \dfrac{|\alpha K_1^* - \hat{\alpha} K_1^*|^2}{|K_1 - \hat{\alpha} \alpha^* K_1|^2} \right) \\
      & = \qty{10}{dB} \cdot \log_{10} \left( \dfrac{|\alpha - \hat{\alpha}|^2}{|1 - \hat{\alpha} \alpha^*|^2} \right) \\
      & \approx \qty{10}{dB} \cdot \log_{10} \left( |\alpha - \hat{\alpha}|^2 \right),
  \end{aligned}
\end{equation}
where the approximation assumes typical direct-conversion receiver implementations where $\alpha \hat{\alpha} \approx 0$.

\section{CRLB Derivation} \label{sec:CRLB}
As will be detailed in the following, the \ac{CRLB} derivation in this work is based on the assumption that each sample of the signal $\ve{r}$ used for estimation follows a general complex Gaussian distribution, even in the absence of any \ac{WGN} component. 

\subsection{Gaussian Assumption}
The Gaussian assumption is based on \ac{CLT} arguments, as the transmitter's \ac{IDFT} performs a linear combination of multiple independent identically distributed random variables $\{d_n[k]\, \forall \,k \in \{0, 1, \dots, \Ndft - 1\}\}$ for each sample of $x[n]$.
While the \ac{PDF} of $x[n]$ approaches a Gaussian distribution for large $L_s$, the approximation degrades for small allocations since fewer random variables contribute to the sum.

\begin{figure}
  \centering
\definecolor{mycolor1}{rgb}{0.00000,0.44700,0.74100}%
\definecolor{mycolor2}{rgb}{0.85000,0.32500,0.09800}%
\definecolor{mycolor3}{rgb}{0.92900,0.69400,0.12500}%
\definecolor{mycolor4}{rgb}{0.49400,0.18400,0.55600}%
\definecolor{mycolor5}{rgb}{0.46600,0.67400,0.18800}%
\begin{tikzpicture}

\newcommand{\XMIN}{1}
\newcommand{\XMAX}{1000}
\newcommand{\NSAMPLES}{1000}
\newcommand{\LINEWIDTH}{1.0pt}

\begin{axis}[%
  xmode=log, 
  log basis x=10, 
  width=.8\linewidth,
  height=.4\linewidth,
  scale only axis,
  xmin=\XMIN,
  xmax=\XMAX,
  xlabel style={font=\color{white!15!black}},
  xlabel={Number of allocated subcarriers $L_s$},
  xticklabel style={font=\normalsize},
  ymin=-1,
  ymax=0.1,
  ylabel style={font=\color{white!15!black}},
  ylabel={Normalized kurtosis $\kappa_4$},
  yticklabel style={font=\normalsize},
  axis background/.style={fill=white},
  xmajorgrids,
  xminorgrids,
  ymajorgrids,
  legend style={
    at={(0.97,0.03)}, 
    anchor=south east, 
    legend cell align=left, 
    align=left, 
    draw=white!15!black,
    font=\normalsize}
]

\addplot[
  domain=\XMIN:\XMAX, 
  samples=\NSAMPLES,
  color=mycolor1,
  line width=\LINEWIDTH]
{
  -3/5*(4+1)/(4-1)/x
}; 
\addlegendentry{$4$-QAM}

\addplot[
  domain=\XMIN:\XMAX, 
  samples=\NSAMPLES,
  color=mycolor2,
  line width=\LINEWIDTH
] 
{
  -3/5*(16+1)/(16-1)/x
}; 
\addlegendentry{$16$-QAM}

\addplot[
  domain=\XMIN:\XMAX, 
  samples=\NSAMPLES,
  color=mycolor3,
  line width=\LINEWIDTH
] 
{
  -3/5*(64+1)/(64-1)/x
}; 
\addlegendentry{$64$-QAM}

\addplot[
  domain=\XMIN:\XMAX, 
  samples=\NSAMPLES,
  color=mycolor4,
  line width=\LINEWIDTH
] 
{
  -3/5*(256+1)/(256-1)/x
}; 
\addlegendentry{$256$-QAM}

\addplot[
  domain=\XMIN:\XMAX, 
  samples=\NSAMPLES,
  color=mycolor5,
  line width=\LINEWIDTH
] 
{
  -3/5*(1024+1)/(1024-1)/x
}; 
\addlegendentry{$1024$-QAM}

\end{axis}
\end{tikzpicture}%
  \caption{Normalized kurtosis versus number of allocated subcarriers $L_s$ with $\Ndft = 4096$ for different QAM modulation orders.}
  \label{fig:kurtosis_over_subcarriers}
\end{figure}
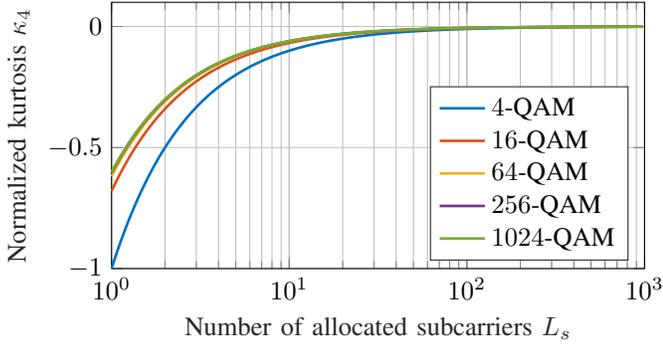

To provide a qualitative indication of the influence of $L_s$ on the Gaussianity of $x[n]$, its normalized kurtosis $\kappa_4$ can be calculated according to \cite{mathis_on-the-kurtosis_2001}. A value of $\kappa_4 = 0$ is a necessary condition for $x[n]$ being Gaussian distributed, whereas $\kappa_4 < 0$ represents a sub-Gaussian distribution \cite{mathis_on-the-kurtosis_2001}. In Fig.~\ref{fig:kurtosis_over_subcarriers}, the normalized kurtosis $\kappa_4$ is plotted against $L_s$ with $\Ndft = 4096$ for different \ac{QAM} orders.
From this plot, it is apparent that the assumption is violated for small $L_s$ (e.g. $L_s < 12$). This limitation should be considered when interpreting subsequent results. Despite this limitation, we still assume a Gaussian distribution for $x[n]$ in the following sections. As will become evident in Section~\ref{sec:simulations}, our derivation still provides a reasonable bound for practical estimator implementations, even in the cases where this Gaussian assumption is clearly violated.

\subsection{Statistical Signal Model} \label{sec:statistical_signal_model}
The derivations in this subsection and Section~\ref{sec:crlb_derivation} closely follow our conference paper~\cite{tockner_crlb_2025}, and are included here to establish the notation and intermediate results required for the novel contributions in Sections~\ref{sec:crlb_optimized} and~\ref{sec:crlb_simplified}.

In this paper, we utilize the complex augmented notation of \cite{adali_complex-valued_2011}, where we refer to the covariance matrix of a complex augmented vector $\ave{x} = [\ve{x}^T, \ve{x}^H]^T$ as
\begin{equation}
  \am{C}_\ve{x} = \Exp{\ave{x} \, \ave{x}^H} = \begin{bmatrix}
    \Exp{\ve{x} \ve{x}^H} & \Exp{\ve{x} \ve{x}^T} \\
    \Exp{\ve{x}^* \ve{x}^H} & \Exp{\ve{x}^* \ve{x}^T}
  \end{bmatrix}
  = \begin{bmatrix}
    \m{C}_\ve{x} & \boldsymbol{\Gamma}_\ve{x} \\
    \boldsymbol{\Gamma}_\ve{x}^* & \m{C}_\ve{x}^*
  \end{bmatrix},
\end{equation}
with $\m{C}_\ve{x}$, and $\boldsymbol{\Gamma}_{\ve{x}}$ as the covariance matrix and the pseudo-covariance matrix of $\ve{x}$, respectively.

Given the \ac{CLT} assumption and the signal model in Section~\ref{sec:signal_model}, $\ave{r}$ is modeled to follow a zero-mean general complex Gaussian \ac{PDF} \cite{adali_complex-valued_2011}
\begin{equation} \label{eq:pdf_r}
  p(\ave{r}; \boldsymbol{\uptheta}) = \dfrac{1}{\pi^N \sqrt{\det \am{C}_{\ve{r}}}} \exp \left(-\dfrac{1}{2} \ave{r}^H \am{C}_{\ve{r}}^{-1} \ave{r} \right),
\end{equation}
where $\text{det}(\cdot)$ denotes the matrix determinant, and $\boldsymbol{\uptheta} = [K_1, K_2]^T$. 
In the following, we derive the $2 N \times 2 N$ block matrix $\am{C}_{\ve{r}}$ by propagating assumptions on $\ve{d}$ (i.e., uniform distribution over the modulation alphabet, statistical independence across subcarriers and \ac{OFDM} symbols, and properness for square \ac{QAM}) through the processing chain in Fig.~\ref{fig:iq_imbalance_bock_diagram}. This expresses $\am{C}_{\ve{r}}$ as a function of $\mathcal{P} = \{K_1, \alpha,\boldsymbol{\psi}, \sigma_{\eta_s}^2, \sigma_{\eta_r}^2, \sigma_d^2, \ve{h}\}$, where $\sigma_d^2$ denotes the average power of $d[k]$, defined as 
\begin{equation} \label{eq:var_d}
  \sigma_{d}^2 = \dfrac{1}{\Ndft} \sum_{k=0}^{\Ndft - 1} \Exp{|d[k]|^2}.
\end{equation}
Due to the focus on blind estimation, $\ve{d}$ is a random vector. From the assumption of uniformly distributed data symbols, the \ac{PMF} of $d[k]$ is defined as
\begin{equation} \label{eq:pmf_d}
  p_{d[k]}(x) =
    \begin{cases}
      \left\{
      \begin{array}{lr}
        \dfrac{1}{|\mathcal{A}|} & \forall x \in \mathcal{A}  \\
        0 & \text{otherwise}
      \end{array} \right\} & \text{for } [\boldsymbol{\psi}]_k \neq 0 \\
      \left\{
      \begin{array}{lr}
        1 & \text{for } x = 0 \\
        0 & \text{otherwise}
      \end{array} \right\} & \text{for } [\boldsymbol{\psi}]_k = 0.
    \end{cases}
\end{equation}
Additionally, we assume statistical independence of the data symbols across subcarriers and \ac{OFDM} symbols and, without loss of generality, a constant mean signal energy per \ac{OFDM} symbol. Thus, the covariance matrix of $\ve{d}_n$ is defined as 
\begin{equation} \label{eq:C_d_n}
  \m{C}_{\ve{d}_n} = \dfrac{\sigma_{d}^2}{L_s} \, \text{diag}(\boldsymbol{\psi}),
\end{equation}
with $\text{diag}(\boldsymbol{\psi})$ representing a matrix with the elements of $\boldsymbol{\psi}$ on its main diagonal, and all other elements being zero.
From \cref{eq:tilde_x,eq:x,eq:C_d_n} we can calculate
\begin{equation} \label{eq:C_x_n}
  \begin{aligned}
    \m{C}_{\ve{x}_n} & = \m{D}_{\text{CP}} \m{F}_{\Ndft}^H \m{C}_{\ve{d}_n} \m{F}_{\Ndft} \m{D}_{\text{CP}}^H,
  \end{aligned}
\end{equation}
since $\m{F}_{\Ndft}^{-1} = \m{F}_{\Ndft}^H$. Given the constant subcarrier allocation assumption, we have $\m{C}_{\ve{d}_n} = \m{C}_{\ve{d}_m}$ and $\m{C}_{\ve{x}_n} = \m{C}_{\ve{x}_m}$ for all $n,m\in \{0, 1, \ldots, N_{\text{OFDM}} - 1\}$. Incorporating the statistical independence between \ac{OFDM} symbols, the extension of the single \ac{OFDM} symbol covariance matrices $\m{C}_{\ve{d}_n}$ and $\m{C}_{\ve{x}_n}$ to the full signal vector covariance matrices $\m{C}_{\ve{d}}$ and $\m{C}_{\ve{x}}$ becomes
\begin{equation}
  \begin{aligned}
    \m{C}_{\ve{d}} & = \m{I}_{N_{\text{OFDM}}} \otimes \m{C}_{\ve{d}_n} \\
    \m{C}_{\ve{x}} & = \m{I}_{N_{\text{OFDM}}} \otimes \m{C}_{\ve{x}_n}    
  \end{aligned} \quad \forall n \in \{0, 1, ..., N_{\text{OFDM}} - 1\}.
\end{equation}
From \eqref{eq:check_s}, it follows that
\begin{equation} \label{eq:C_check_s}
  \m{C}_{\check{\ve{s}}} = \m{H} \m{C}_{\ve{x}} \m{H}^H.
\end{equation}
Including the \ac{AWGN} from \eqref{eq:s} leads to 
\begin{equation} \label{eq:C_s}
  \m{C}_{\ve{s}} = \m{C}_{\check{\ve{s}}} + \sigma_{\eta_s}^2 \m{I}_{N}.
\end{equation}
It is known from \cite{paireder_enhanced_2021,anttila_circularity-based_2008,anttila_circularity_2011} that the pseudo-covariance matrix $\m{\Gamma}_{\ve{d}} = \m{0}_{N \times N}$ for \ac{QPSK} and square M-\ac{QAM}. Moreover, the processing steps in \cref{eq:tilde_x,eq:x,eq:check_s,eq:s} are all linear. Therefore, $\m{\Gamma}_{\ve{d}} = \m{0}_{N \times N}$ implies $\m{\Gamma}_{\ve{s}} = \m{0}_{N \times N}$.
Consequently, 
\begin{equation} \label{eq:C_s_aug}
  \am{C}_\ve{s} = \begin{bmatrix}
    \m{C}_{\ve{s}} & \m{0}_{N \times N} \\
    \m{0}_{N \times N} & \m{C}_{\ve{s}}^*
  \end{bmatrix},
\end{equation}
and from \cref{eq:K,eq:check_aug_r,eq:C_s_aug} it follows that
\begin{equation} \label{eq:C_check_r_aug}
  \begin{aligned}
      \am{C}_{\check{\ve{r}}} & = \am{K} \, \am{C}_{\ve{s}} \, \am{K}^H \\
      & = \begin{bmatrix}
          |K_1|^2 \m{C}_{\ve{s}} + |K_2|^2 \m{C}_{\ve{s}}^* & 2 K_1 K_2 \Re\{\m{C}_{\ve{s}}\} \\
          2 K_1^* K_2^* \Re\{\m{C}_{\ve{s}}\} & |K_1|^2 \m{C}_{\ve{s}}^* + |K_2|^2 \m{C}_{\ve{s}}
      \end{bmatrix}.
  \end{aligned}
\end{equation}
Finally, the required $2 N \times 2 N$ augmented covariance matrix of $\ve{r}$ is given by
\begin{equation} \label{eq:C_r_aug}
  \begin{aligned}
    \am{C}_{\ve{r}} & = \am{C}_{\check{\ve{r}}} + \sigma_{\eta_r}^2 \m{I}_{2 N} \\
    & = \begin{bmatrix}
      \m{C}_{\check{\ve{r}}} + \sigma^2_{\eta_r} \m{I}_{N} & \m{\Gamma}_{\check{\ve{r}}} \\
      \m{\Gamma}_{\check{\ve{r}}}^* & \m{C}_{\check{\ve{r}}}^* + \sigma^2_{\eta_r} \m{I}_{N}
    \end{bmatrix},
  \end{aligned}
\end{equation}
with $\m{C}_{\check{\ve{r}}}$ and $\m{\Gamma}_{\check{\ve{r}}}$ as the northwest and northeast blocks of $\am{C}_{\check{\ve{r}}}$, respectively.

\subsection{Cram\'{e}r-Rao Lower Bound} \label{sec:crlb_derivation}
Let 
\begin{equation}
  \am{J}_F = \begin{bmatrix}
    \m{J}_F & \tilde{\m{J}}_F \\
    \tilde{\m{J}}_F^* & \m{J}_F^*
  \end{bmatrix},
\end{equation}
be the complex augmented \ac{FIM}, with $\m{J}_F$ as the classical \ac{FIM} and $\tilde{\m{J}}_F$ as the pseudo-\ac{FIM}. Then, if $\am{J}_F$ is invertible, the widely linear \ac{CRLB} for estimating complex-valued parameters from a complex-valued signal corresponds to the northwest block of the inverse complex augmented \ac{FIM} $\am{J}_F^{-1}$ \cite{schreier_statistical_2010}.
%In general, the widely linear \ac{CRLB} for estimating complex-valued parameters from a complex-valued signal corresponds to the northwest block of the inverse complex augmented \ac{FIM} $\am{J}_F^{-1}$ \cite{schreier_statistical_2010}. 
Due to its block-matrix structure, this northwest block is expressed by $[\m{J}_F - \tilde{\m{J}}_F \m{J}_F^{-*} \tilde{\m{J}}_F^*]^{-1}$. For any unbiased estimator the covariance matrix $\m{C}_{\hat{\boldsymbol{\uptheta}}}$ is then bounded by \cite{schreier_statistical_2010}
\begin{equation} \label{eq:C_theta_hat_aug}
  \m{C}_{\hat{\boldsymbol{\uptheta}}} \geq [\m{J}_F - \tilde{\m{J}}_F \m{J}_F^{-*} \tilde{\m{J}}_F^*]^{-1}.
\end{equation}
However, for the estimation problem considered here, $\am{J}_F$ is singular, implying that the inverse in \eqref{eq:C_theta_hat_aug} does not exist. This singularity does not arise from the trivial case $\m{J}_F = \tilde{\m{J}}_F$ discussed in \cite{schreier_statistical_2010}, but rather from the fact that the two complex-valued parameters $K_1$ and $K_2$ are dependent. Both are functions of the two real-valued \ac{I/Q} imbalance parameters $\epsilon$ and $\phi$. Consequently, it suffices to consider the classical $2 \times 2$ \ac{FIM} $\m{J}_F$ for the improper multivariate Gaussian model, defined as \cite{schreier_statistical_2010}
\begin{equation}
  \label{eq:J_F}
  [\m{J}_F]_{k,l} = \dfrac{1}{2} \tr*{\am{C}_{\ve{r}}^{-1} \dfrac{\partial \am{C}_{\ve{r}}}{\partial \theta_k} \am{C}_{\ve{r}}^{-1} \dfrac{\partial \am{C}_{\ve{r}}}{\partial \theta_l^*} },
\end{equation}
where $\tr{\cdot}$ denotes the trace operator.

In order to calculate the bound for estimating the more commonly used \ac{I/Q} imbalance parameter $\alpha$, $K_2$ can be substituted by $\alpha K_1^*$ in \cref{eq:C_check_r_aug}, leading to
\begin{equation} \label{eq:C_check_r_aug_new}
  \am{C}_{\check{\ve{r}}} 
  = |K_1|^2 \begin{bmatrix}
      \m{C}_{\ve{s}} + |\alpha|^2 \m{C}_{\ve{s}}^* & 2 \alpha \Re\{\m{C}_{\ve{s}}\} \\
      2 \alpha^* \Re\{\m{C}_{\ve{s}}\} & \m{C}_{\ve{s}}^* + |\alpha|^2 \m{C}_{\ve{s}}
  \end{bmatrix}.
\end{equation}
Consequently, we define a new parameter vector $\boldsymbol{\uptheta}_{\alpha} = [K_1, \alpha]^T$. 
Using Wirtinger calculus \cite{wirtinger_zur-formalen_1927}, the partial derivatives in \cref{eq:J_F} (with respect to the elements of $\boldsymbol{\uptheta}_{\alpha}$ and $\boldsymbol{\uptheta}_{\alpha}^*$ instead of $\boldsymbol{\uptheta}$ and $\boldsymbol{\uptheta}^*$) are then calculated from \cref{eq:C_r_aug,eq:C_check_r_aug_new,eq:C_s} as
\begin{align}
  \label{eq:d_aug_C_r_d_K_1_with_alpha}
  \dfrac{\partial \am{C}_{\ve{r}}}{\partial K_1} & = K_1^* \begin{bmatrix}
    \m{C}_{\ve{s}} + |\alpha|^2 \m{C}_{\ve{s}}^* & 2 \alpha \Re\{\m{C}_{\ve{s}}\} \\
    2 \alpha^* \Re\{\m{C}_{\ve{s}}\} & |\alpha|^2 \m{C}_{\ve{s}} + \m{C}_{\ve{s}}^*
  \end{bmatrix}, \\
  \label{eq:d_aug_C_r_d_K_1_conj_with_alpha}
  \dfrac{\partial \am{C}_{\ve{r}}}{\partial K_1^*} & = \dfrac{K_1}{K_1^*} \dfrac{\partial \am{C}_{\ve{r}}}{\partial K_1}, \\
  \label{eq:d_aug_C_r_d_alpha}
  \dfrac{\partial \am{C}_{\ve{r}}}{\partial \alpha} & = |K_1|^2 \begin{bmatrix}
    \alpha^* \m{C}_{\ve{s}}^* & 2 \Re\{\m{C}_{\ve{s}}\} \\
    \m{0}_{N \times N} & \alpha^* \m{C}_{\ve{s}}
  \end{bmatrix}, \\
  \label{eq:d_aug_C_r_d_alpha_conj}
  \dfrac{\partial \am{C}_{\ve{r}}}{\partial \alpha^*} & = |K_1|^2 \begin{bmatrix}
    \alpha \m{C}_{\ve{s}}^* & \m{0}_{N \times N} \\
    2 \Re\{\m{C}_{\ve{s}}\} & \alpha \m{C}_{\ve{s}}
  \end{bmatrix}. 
\end{align}
Finally, \cref{eq:J_F,eq:d_aug_C_r_d_K_1_with_alpha,eq:d_aug_C_r_d_K_1_conj_with_alpha,eq:d_aug_C_r_d_alpha,eq:d_aug_C_r_d_alpha_conj} allow calculating the \ac{CRLB} for jointly estimating $K_1$ and $\alpha$ as a function of the parameters in $\mathcal{P}$. The \ac{CRLB} for estimating $\alpha$ alone, treating $K_1$ as a nuisance parameter, is
\begin{equation} \label{eq:Var_alpha}
    \Var{\hat{\alpha}} = [\m{C}_{\hat{\boldsymbol{\uptheta}}_{\alpha}}]_{2,2} \geq [\m{J}_F^{-1}]_{2,2} = \frac{[\m{J}_F]_{1,1}}{[\m{J}_F]_{1,1}[\m{J}_F]_{2,2} - |[\m{J}_F]_{1,2}|^2},
\end{equation}
with $\m{J}_F$ from \eqref{eq:J_F} where $\theta_1 = K_1$ and $\theta_2 = \alpha$.

\subsection{Optimized Calculation} \label{sec:crlb_optimized}
Inverting the $2N \times 2N$ augmented covariance matrix $\am{C}_{\ve{r}}$ entails a high computational complexity of $\mathcal{O}(N^3)$, and might also become quite memory-intensive for reasonably sized $N_{\text{OFDM}}$ and $\Nsym$. This motivates the need for an optimized calculation of the \ac{CRLB}. 
The first optimization is an approximation based on findings from simulation results. We observe that the \ac{CRLB} is independent of $L_{\text{cp}}$ when $\sigma_{\eta_s}^2 = 0$. For $\sigma_{\eta_s}^2 > 0$, comparing the bound for a typical ratio of $L_{\text{cp}} / \Ndft$ with $L_{\text{cp}} / \Ndft = 0$ reveals a negligible influence of $L_{\text{cp}}$. 
%An intuitive explanation of these findings is that the copied samples in the \ac{CP} do not provide additional information about the \ac{I/Q} imbalance. Only the additional \ac{I/Q} imbalanced noise samples contribute information. However, the combination of realistically small $L_{\text{cp}} / \Ndft$ and $\sigma_{\eta_s}^2$ results in very little additional information, and therefore also a negligible impact on the \ac{CRLB}. 
Based on these findings, we omit the rows and columns in the covariance matrices corresponding to the \ac{CP}. Consequently, $\m{C}_{\check{\ve{s}}}$ shares the block-diagonal structure of $\m{C}_{\ve{x}}$, as do $\m{\Gamma}_{\check{\ve{s}}}$, $\m{C}_{\ve{s}}$, $\m{\Gamma}_{\check{\ve{r}}}$, $\m{C}_{\check{\ve{r}}}$, $\m{\Gamma}_{\ve{r}}$, and $\m{C}_{\ve{r}}$. 

Furthermore, this block-diagonal structure of all matrices involved in the trace operator of \cref{eq:J_F} enables an additional optimization. The trace can be evaluated simply by computing $\am{C}_{\ve{r}}^{-1} (\partial \am{C}_{\ve{r}} / \partial \theta_k) \am{C}_{\ve{r}}^{-1} (\partial \am{C}_{\ve{r}} / \partial \theta_l^*)$ for a single \ac{OFDM} symbol (i.e., $\am{C}_{\ve{r}} \in \mathbb{C}^{2 \Ndft \times 2 \Ndft}$) and scaling the result by $N_{\text{OFDM}}$. Consequently, the \ac{FIM} entries are calculated as \cite{schreier_statistical_2010}
\begin{equation}
  \label{eq:J_F_simple}
  [\m{J}'_F]_{k,l} = \dfrac{N_{\text{OFDM}}}{2} \tr*{\am{C}_{\ve{r}}'^{-1} \dfrac{\partial \am{C}_{\ve{r}}'}{\partial \theta_k} \am{C}_{\ve{r}}'^{-1} \dfrac{\partial \am{C}_{\ve{r}}'}{\partial \theta_l^*} },
\end{equation}
with 
\begin{equation}
    \am{C}_{\ve{r}}' = \begin{bmatrix}
      \m{C}_{\ve{r}}' & \m{\Gamma}_{\ve{r}}' \\
      \m{\Gamma}_{\ve{r}}'^* & \m{C}_{\ve{r}}'^*
    \end{bmatrix},
\end{equation} 
where $\m{C}_{\ve{r}}' = \m{C}_{\ve{r}}[\mathcal{I}]$ and $\m{\Gamma}_{\ve{r}}' = \m{\Gamma}_{\ve{r}}[\mathcal{I}]$ denote principal submatrices \cite{horn_matrix_2012} with index set $\mathcal{I} = \{L_{\text{cp}}+1, \ldots, \Nsym\}$.
This reduces the computational complexity from $\mathcal{O}(N^3)$ to $\mathcal{O}(N^3_{\textrm{OFDM}})$, since the matrix to be inverted is now of dimension $2\Ndft \times 2\Ndft$.

The next optimization exploits the property that the trace of a matrix equals the sum of its eigenvalues. Given two matrices $\m{A} = \m{V} \m{\Lambda}_{\m{A}} \m{V}^{-1}$ and $\m{B} = \m{V} \m{\Lambda}_{\m{B}} \m{V}^{-1}$ sharing a common eigenvector matrix $\m{V}$, the following identities hold:
\begin{equation} \label{eq:eigendecomposition_trace}
  \begin{aligned}
    \m{A}^{-1} & = \m{V} \m{\Lambda}_{\m{A}}^{-1} \m{V}^{-1}, \\
    \m{A} + \m{B} & = \m{V} \left(\m{\Lambda}_{\m{A}} + \m{\Lambda}_{\m{B}}\right) \m{V}^{-1}, \\
    \m{A} \m{B} & = \m{V} \m{\Lambda}_{\m{A}} \m{\Lambda}_{\m{B}} \m{V}^{-1}, \\
    \m{A} + a \m{I} & = \m{V} \left(\m{\Lambda}_{\m{A}} + a \m{I}\right) \m{V}^{-1},
  \end{aligned}
\end{equation}
where $\m{\Lambda}_{\m{A}}$ and $\m{\Lambda}_{\m{B}}$ are diagonal matrices containing the eigenvalues of $\m{A}$ and $\m{B}$, respectively. These identities allow expressing the product inside the trace operator in \cref{eq:J_F_simple} solely in terms of the eigenvalues of $\m{C}_{\ve{r}}'$ and $\m{\Gamma}_{\ve{r}}'$. In the following, we show how to obtain these eigenvalues directly from the known $\boldsymbol{\psi}$ and $\sigma_d^2$, without performing an explicit eigenvalue decomposition. This reduces the computational complexity even further to $\mathcal{O}(\Ndft)$.

Analogous to the covariance propagation in Section~\ref{sec:statistical_signal_model}, we now derive the eigenvalues of the CP-removed covariance matrices $\m{C}'_{\ve{r}}$ and $\m{\Gamma}'_{\ve{r}}$, beginning with the transmitted signal covariance matrix
\begin{equation}
  \m{C}_{\ve{x}}' = \m{C}_{\ve{x}}[\mathcal{I}] = \m{F}_{\Ndft}^H \m{C}_{\ve{d}} \m{F}_{\Ndft}.
\end{equation}
From this representation, it follows that $\m{C}_{\ve{d}}$ is the eigenvalue matrix of $\m{C}_{\ve{x}}'$.
Deriving $\m{C}_{\check{\ve{s}}}'$ begins with the time-domain signal relation 
\begin{equation}
   \check{\ve{s}}'_n = \m{H}' \ve{x}'_n = \m{H}' \m{F}_{\Ndft}^H \ve{d}_n,   
\end{equation}
where $\check{\ve{s}}'_n = \check{\ve{s}}_n[\mathcal{I}]$ and $\ve{x}'_n = \ve{x}_n[\mathcal{I}]$ denote the signal vectors excluding the \ac{CP}, with the submatrix notation naturally extended to subvectors, and $\m{H}' \in \mathbb{C}^{\Ndft \times \Ndft}$ is the circular-convolution channel matrix
\begin{multline}
  \m{H}' = \\ \scalebox{0.9}{$\begin{bmatrix}
    h[0] & 0 & \hdots & h[Q{-}1] & \hdots & h[1] \\
    h[1] & h[0] & & & \ddots & \vdots \\
    \vdots & \ddots & \ddots & & & h[Q{-}1] \\
    h[Q{-}1] & & \ddots & \ddots & & 0 \\
    0 & h[Q{-}1] & & \ddots & \ddots & \vdots \\
    \vdots & \ddots & \ddots & \hdots & h[0] & 0 \\
    0 & \hdots & h[Q{-}1] & \hdots & h[1] & h[0]
  \end{bmatrix}$}.
\end{multline}
The circular structure arises because the \ac{CP} absorbs the linear convolution transient (for $\Lcp > Q$), effectively converting the linear convolution with $\m{H}$ into a circular convolution after \ac{CP} removal.
In the frequency domain, this yields 
\begin{equation}
  \check{\ve{s}}'_{n,f} = \m{F}_{\Ndft} \check{\ve{s}}'_{n} = \m{H}_f \ve{d}_n
\end{equation}
with the diagonal channel transfer function matrix 
\begin{equation}
  \m{H}_f = \m{F}_{\Ndft} \m{H}' \m{F}_{\Ndft}^H = \text{diag}(\m{F}_{\Ndft}[\varnothing^c, \mathcal{J}] \, \ve{h}),
\end{equation}
where $\varnothing^c$ denotes the complement of the empty set (i.e., all row indices) and $\mathcal{J} = \{1, \ldots, Q\}$ \cite{horn_matrix_2012}.
Consequently, the covariance matrix of $\check{\ve{s}}'_{n}$ is given by
\begin{equation} \label{eq:C_check_s_prime}
  \begin{aligned}
    \m{C}'_{\check{\ve{s}}} & = \m{C}_{\check{\ve{s}}}[\mathcal{I}] \\
    & = \m{H}' \m{C}'_{\ve{x}} \m{H}'^H \\
    & = \m{H}' \m{F}_{\Ndft}^H \m{C}_{\ve{d}} \m{F}_{\Ndft} \m{H}'^H \\
    & = \m{F}_{\Ndft}^H \m{F}_{\Ndft} \m{H}' \m{F}_{\Ndft}^H \m{C}_{\ve{d}} \m{F}_{\Ndft} \m{H}'^H \m{F}_{\Ndft}^H \m{F}_{\Ndft} \\
    & = \m{F}_{\Ndft}^H \m{H}_{f} \m{C}_{\ve{d}} \m{H}_{f}^H \m{F}_{\Ndft} \\
    & = \m{F}_{\Ndft}^H \m{C}_{\check{\ve{s}},f}' \m{F}_{\Ndft},
  \end{aligned}
\end{equation}
with the diagonal channel-distorted frequency-domain signal covariance matrix 
\begin{equation}
  \m{C}_{\check{\ve{s}},f}' = \m{H}_{f} \m{C}_{\ve{d}} \m{H}_{f}^H.
\end{equation}
Since $\m{C}_{\ve{r}}'$ is constructed from $\m{C}_{\ve{s}}'$ and $\m{C}_{\ve{s}}'^*$, we also require the eigenstructure of the conjugate complex covariance matrix. To this end, we note that
\begin{equation}
  \begin{aligned}
    \m{C}_{\check{\ve{s}}}'^* & = (\m{F}_{\Ndft}^H \m{C}_{\check{\ve{s}},f}' \m{F}_{\Ndft})^* \\
    & = \m{F}_{\Ndft} \m{C}_{\check{\ve{s}},f}' \m{F}_{\Ndft}^H \\
    & = \m{F}_{\Ndft}^H \m{M} \m{C}_{\check{\ve{s}},f}' \m{M} \m{F}_{\Ndft},
  \end{aligned}
\end{equation}
introducing the mirroring matrix
\begin{equation}
  \m{M} = \m{F}_{\Ndft}^2 = \begin{bmatrix}
    1 & 0 & \cdots & 0 \\
    0 & 0 & \cdots & 1 \\
    \vdots & \vdots & \iddots & \vdots \\
    0 & 1 & \cdots & 0
  \end{bmatrix}.
\end{equation}
The left- and right-hand side multiplication of $\m{C}_{\check{\ve{s}},f}'$ with $\m{M}$ leads to 
\begin{multline}
  [\m{M} \m{C}_{\check{\ve{s}},f}' \m{M}]_{n,n} \\ = \begin{cases}
    [\m{C}_{\check{\ve{s}},f}']_{n,n} & \text{for } n = 1 \\
    [\m{C}_{\check{\ve{s}},f}']_{\Ndft - n + 2,\Ndft - n + 2} & \text{for } n = 2, ..., \Ndft, \\
  \end{cases}
\end{multline}
representing a swap of the diagonal elements with their subcarrier-respective image. Therefore, we denote
\begin{equation} \label{eq:C_check_s_f_prime_img}
  \m{C}_{\check{\ve{s}},f,\text{Img}}' = \m{M} \m{C}_{\check{\ve{s}},f}' \m{M},
\end{equation}
where the subscript ``Img'' indicates left- and right-hand side multiplication by $\m{M}$. This notation is used consistently throughout the remainder of this paper. Accordingly,
\begin{equation} \label{eq:C_check_s_prime_conj}
  \m{C}_{\check{\ve{s}}}'^* = \m{F}_{\Ndft}^H \m{C}_{\check{\ve{s}},f,\text{Img}}' \m{F}_{\Ndft}.
\end{equation}
Substituting \cref{eq:C_check_s_prime} into the \ac{CP}-dropped analog of \cref{eq:C_s} yields
\begin{equation}
  \begin{aligned}
    \m{C}_{\ve{s}}' & = \m{F}_{\Ndft}^H \m{C}_{\check{\ve{s}},f}' \m{F}_{\Ndft} + \sigma_{\eta_s}^2 \m{I}_{\Ndft} \\
    & = \m{F}_{\Ndft}^H \left(\m{C}_{\check{\ve{s}},f}' + \sigma_{\eta_s}^2 \m{I}_{\Ndft}\right) \m{F}_{\Ndft},
  \end{aligned}
\end{equation}
and similarly for \cref{eq:C_check_s_prime_conj},
\begin{equation}
  \m{C}_{\ve{s}}'^* = \m{F}_{\Ndft}^H \left(\m{C}_{\check{\ve{s}},f,\text{Img}}' + \sigma_{\eta_s}^2 \m{I}_{\Ndft}\right) \m{F}_{\Ndft}.
\end{equation}
Inserting these frequency-domain representations of $\m{C}_{\ve{s}}'$ and $\m{C}_{\ve{s}}'^*$ into the \ac{CP}-dropped analog of \cref{eq:C_check_r_aug_new}, and using $2\Re\{\m{C}_{\ve{s}}'\} = \m{C}_{\ve{s}}' + \m{C}_{\ve{s}}'^*$ for the off-diagonal blocks, we obtain
\begin{align}
  \label{eq:C_check_r_prime}
  \m{C}_{\check{\ve{r}}}' & = \m{F}_{\Ndft}^H |K_1|^2 \left(
    \m{C}_{\ve{s},f}' + |\alpha|^2  \m{C}_{\ve{s},f,\text{Img}}'
    \right) \m{F}_{\Ndft}, \\
  \label{eq:Gamma_check_r_prime}
  \m{\Gamma}_{\check{\ve{r}}}' & = \m{F}_{\Ndft}^H \alpha |K_1|^2 \left( 
    \m{C}_{\ve{s},f}' + \m{C}_{\ve{s},f,\text{Img}}'
  \right) \m{F}_{\Ndft},
\end{align}
where
\begin{align}
  \m{C}_{\ve{s},f}' & = \m{C}_{\check{\ve{s}},f}' + \sigma_{\eta_s}^2 \m{I}_{\Ndft} \\
  \m{C}_{\ve{s},f,\text{Img}}' & = \m{C}_{\check{\ve{s}},f,\text{Img}}' + \sigma_{\eta_s}^2 \m{I}_{\Ndft}.
\end{align}
Substituting \cref{eq:C_check_r_prime,eq:Gamma_check_r_prime} into the \ac{CP}-dropped analog of \cref{eq:C_r_aug} leads to 
\begin{align}
  \label{eq:C_r_prime}
  \m{C}_{\ve{r}}' & = \m{F}_{\Ndft}^H \left(\m{C}_{\check{\ve{r}},f}' + \sigma_{\eta_r}^2 \m{I}_{\Ndft} \right) \m{F}_{\Ndft}, \\
  \label{eq:Gamma_r_prime}
  \m{\Gamma}_{\ve{r}}' & = \m{\Gamma}_{\check{\ve{r}}}' = \m{F}_{\Ndft}^H \m{\Gamma}_{\check{\ve{r}},f}' \m{F}_{\Ndft},
\end{align}
where
\begin{align}
  \m{C}_{\check{\ve{r}},f}' & = |K_1|^2 \left(
          \m{C}_{\ve{s},f}' + |\alpha|^2  \m{C}_{\ve{s},f,\text{Img}}' 
        \right), \\
  \m{\Gamma}_{\check{\ve{r}},f}' & = \alpha |K_1|^2 \left( 
    \m{C}_{\ve{s},f}' + \m{C}_{\ve{s},f,\text{Img}}'
  \right),
\end{align}
from which we additionally obtain 
\begin{align}
  \m{C}_{\ve{r},f}' & = \m{C}_{\check{\ve{r}},f}' + \sigma_{\eta_r}^2 \m{I}_{\Ndft}, \\
  \m{\Gamma}_{\ve{r},f}' & = \m{\Gamma}_{\check{\ve{r}},f}'.
\end{align}

To evaluate the \ac{FIM} in \cref{eq:J_F_simple}, we require the inverse of the augmented covariance matrix $\am{C}_{\ve{r}}'$. Since this matrix has a $2 \times 2$ block structure, the block-matrix inversion lemma \cite{bernstein_matrix_2005} yields
\begin{equation} \label{eq:C_r_prime_inv}
  \am{C}_{\ve{r}}'^{-1} = \begin{bmatrix}
    \m{C}_1^{-1} & - \m{C}_{\ve{r}}'^{-1} \m{\Gamma}_{\ve{r}}' \m{C}_2^{-1} \\
    - \m{C}_2^{-1} \m{\Gamma}_{\ve{r}}'^* \m{C}_{\ve{r}}'^{-1} & \m{C}_2^{-1}
  \end{bmatrix},
\end{equation}
with 
\begin{align}
  \label{eq:C_1}
  \m{C}_1 & = \m{C}_{\ve{r}}' - \m{\Gamma}_{\ve{r}}' \m{C}_{\ve{r}}'^{-*} \m{\Gamma}_{\ve{r}}'^*, \\
  \label{eq:C_2}
  \m{C}_2 & = \m{C}_{\ve{r}}'^* - \m{\Gamma}_{\ve{r}}'^* \m{C}_{\ve{r}}'^{-1} \m{\Gamma}_{\ve{r}}'.
\end{align}
Since $\m{C}_{\ve{r}}'$ and $\m{\Gamma}_{\ve{r}}'$ are diagonalized by $\m{F}_{\Ndft}$, the identities in \cref{eq:eigendecomposition_trace} allow computing the required matrix inverses by element-wise inversion of the eigenvalues. Applying these identities, the blocks of $\am{C}_{\ve{r}}'^{-1}$ in \cref{eq:C_r_prime_inv} simplify to
\begin{align}
  \label{eq:C_r_aug_prime_inv_NW}
  \m{C}_1^{-1} & = \m{F}_{\Ndft}^H \Biggl(     
    \m{C}_{\ve{r},f}' - 
      \bigl| \m{\Gamma}_{\ve{r},f}' \bigr|^2 \m{C}_{\ve{r},f,\text{Img}}'^{-1}
  \Biggr)^{-1} \m{F}_{\Ndft}, \\
  \label{eq:C_r_aug_prime_inv_SE}
  \m{C}_2^{-1} & = \m{F}_{\Ndft}^H \Biggl(     
    \m{C}_{\ve{r},f,\text{Img}}' - 
      \bigl| \m{\Gamma}_{\ve{r},f}' \bigr|^2 \m{C}_{\ve{r},f}'^{-1}
  \Biggr)^{-1} \m{F}_{\Ndft},
\end{align}
and to 
\begin{multline}
  \label{eq:C_r_aug_prime_inv_NE}
  \m{C}_{\ve{r}}'^{-1} \m{\Gamma}_{\ve{r}}' \m{C}_2^{-1} = \\ \m{F}_{\Ndft}^H 
  \left( 
    \m{\Gamma}_{\ve{r},f}' (\m{C}_{\ve{r},f}' \m{C}_{\ve{r},f,\text{Img}}' - |\m{\Gamma}_{\ve{r},f}'|^2)^{-1}
  \right) \m{F}_{\Ndft}
\end{multline}
\begin{multline}
  \label{eq:C_r_aug_prime_inv_SW}
  \m{C}_2^{-1} \m{\Gamma}_{\ve{r}}'^* \m{C}_{\ve{r}}'^{-1} = \\ \m{F}_{\Ndft}^H 
  \left( 
    \m{\Gamma}_{\ve{r},f}'^* (\m{C}_{\ve{r},f}' \m{C}_{\ve{r},f,\text{Img}}' - |\m{\Gamma}_{\ve{r},f}'|^2)^{-1}
  \right) \m{F}_{\Ndft}.
\end{multline}
The \ac{FIM} in \cref{eq:J_F_simple} also requires the derivatives of $\am{C}_{\ve{r}}'$. These take the same form as the unprimed derivatives in \cref{eq:d_aug_C_r_d_K_1_with_alpha,eq:d_aug_C_r_d_K_1_conj_with_alpha,eq:d_aug_C_r_d_alpha,eq:d_aug_C_r_d_alpha_conj}, but with the substitutions $\m{C}_{\check{\ve{s}}} \rightarrow \m{C}_{\check{\ve{s}}}'$ and $\m{C}_{\check{\ve{s}}}^* \rightarrow \m{C}_{\check{\ve{s}}}'^*$. 

Therefore, all factors of the matrix product inside the trace operator of \cref{eq:J_F_simple} can now be formulated as $2 \Ndft \times 2 \Ndft$ block matrices with $\Ndft \times \Ndft$ blocks. These block matrices can be multiplied according to
\begin{equation}  \label{eq:block_matrix_mult}
  \begin{aligned}
    \m{A} \m{B} & =
    \begin{bmatrix}
      \m{A}_{11} & \m{A}_{12} \\
      \m{A}_{21} & \m{A}_{22}
    \end{bmatrix}
    \begin{bmatrix}
      \m{B}_{11} & \m{B}_{12} \\
      \m{B}_{21} & \m{B}_{22}
    \end{bmatrix} \\
    & =
    \begin{bmatrix}
      \m{A}_{11} \m{B}_{11} + \m{A}_{12} \m{B}_{21} & \m{A}_{11} \m{B}_{12} + \m{A}_{12} \m{B}_{22} \\
      \m{A}_{21} \m{B}_{11} + \m{A}_{22} \m{B}_{21} & \m{A}_{21} \m{B}_{12} + \m{A}_{22} \m{B}_{22}
    \end{bmatrix}.
  \end{aligned}
\end{equation}
The key insight for evaluating \cref{eq:J_F_simple} efficiently is that all $\Ndft \times \Ndft$ blocks, including the inverse covariance blocks from \cref{eq:C_r_aug_prime_inv_NW,eq:C_r_aug_prime_inv_NE,eq:C_r_aug_prime_inv_SW,eq:C_r_aug_prime_inv_SE} and the derivative blocks, share the common eigenvector matrix $\m{F}_{\Ndft}$. As a result, products of these blocks are also diagonalized by $\m{F}_{\Ndft}$, with eigenvalues given by element-wise products. Combining this with \cref{eq:eigendecomposition_trace,eq:block_matrix_mult}, the trace of such a product reduces to the sum of eigenvalue products, which reduces the computational complexity from $\mathcal{O}(\Ndft^3)$ to $\mathcal{O}(\Ndft)$.

While the computation is now efficient, writing out the explicit scalar expression for each \ac{FIM} element remains unwieldy: each element involves the trace of a product of four $2 \times 2$ block matrices (two inverse covariance matrices and two derivatives), where each block itself contributes multiple terms through \cref{eq:block_matrix_mult}. The resulting expressions, though straightforward to implement, do not lend themselves to analytical interpretation. To obtain closed-form results that provide insight into how system parameters affect the \ac{CRLB}, the following section introduces a simplifying approximation.

\subsection{Small I/Q Imbalance Approximation} \label{sec:crlb_simplified}
A clearer understanding of the \ac{CRLB} is obtained by combining the optimized calculations with a small \ac{I/Q} imbalance approximation, i.e., $|\alpha| \rightarrow 0$ and $|K_1| \rightarrow 1$. Since the augmented covariance matrix $\am{C}_{\ve{r}}$ and its derivatives depend on $K_1$ and $\alpha$ (cf.\ \cref{eq:C_r_aug}), the \ac{CRLB} in \cref{eq:Var_alpha} is itself a function of the parameters to be estimated. The approximation therefore evaluates this parameter-dependent bound at the operating point $|\alpha| \rightarrow 0$, $|K_1| \rightarrow 1$. This assumption is useful in practice because direct-conversion receiver frontends typically exhibit an $\text{ILR}_{\unit{dB}}$ between \qty{-20}{dB} and \qty{-40}{dB} \cite{paireder_ultra-low_2019}. Furthermore, the simulations presented in Section~\ref{sec:simulations} confirm that the error between the approximated and the actual \ac{CRLB} is negligible, even for the practical worst case of an $\text{ILR}_{\unit{dB}} = \qty{-20}{dB}$.

Under this approximation, the augmented covariance matrix $\am{C}_{\ve{r}}'$ becomes block-diagonal, as the off-diagonal block $\m{\Gamma}_{\ve{r}}'$ (proportional to $\alpha$) vanishes:
\begin{equation}
  \am{C}_{\ve{r}}' \approx \begin{bmatrix} \m{C}_{\ve{r}}' & \m{0} \\ \m{0} & \m{C}_{\ve{r}}'^* \end{bmatrix}.
\end{equation}
Applying the approximation to the derivatives in \cref{eq:d_aug_C_r_d_K_1_with_alpha,eq:d_aug_C_r_d_K_1_conj_with_alpha,eq:d_aug_C_r_d_alpha,eq:d_aug_C_r_d_alpha_conj} yields
\begin{align}
  \label{eq:d_aug_C_r_d_K_1_approx}
  \frac{\partial \am{C}_{\ve{r}}'}{\partial K_1} & \approx \frac{\partial \am{C}_{\ve{r}}'}{\partial K_1^*} \approx \begin{bmatrix} \m{C}_{\ve{s}}' & \m{0} \\ \m{0} & \m{C}_{\ve{s}}'^* \end{bmatrix}, \\
  \label{eq:d_aug_C_r_d_alpha_approx}
  \frac{\partial \am{C}_{\ve{r}}'}{\partial \alpha} & \approx \begin{bmatrix} \m{0} & 2 \Re\{\m{C}_{\ve{s}}'\} \\ \m{0} & \m{0} \end{bmatrix}, \\
  \label{eq:d_aug_C_r_d_alpha_conj_approx}
  \frac{\partial \am{C}_{\ve{r}}'}{\partial \alpha^*} & \approx \begin{bmatrix} \m{0} & \m{0} \\ 2 \Re\{\m{C}_{\ve{s}}'\} & \m{0} \end{bmatrix}.
\end{align}
Furthermore, applying the approximation $\m{\Gamma}_{\ve{r}}' \approx \m{0}$ to \cref{eq:C_1,eq:C_2} results in $\m{C}_1 \approx \m{C}_{\ve{r}}'$ and $\m{C}_2 \approx \m{C}_{\ve{r}}'^*$. Substituting these into \cref{eq:C_r_prime_inv} yields a block-diagonal inverse covariance matrix
\begin{equation} \label{eq:aug_C_r_prime_inv_approx}
  \am{C}_{\ve{r}}'^{-1} \approx \begin{bmatrix} \m{C}_{\ve{r}}'^{-1} & \m{0} \\ \m{0} & \m{C}_{\ve{r}}'^{-*} \end{bmatrix}.
\end{equation}
This structural difference leads to a decoupling of the parameters. Specifically, the cross-terms of the \ac{FIM} ($[\m{J}'_F]_{1,2}$ and $[\m{J}'_F]_{2,1}$) involve the trace of a product of block-diagonal and off-diagonal matrices.
Using \cref{eq:J_F_simple} and the approximated matrices \cref{eq:d_aug_C_r_d_K_1_approx,eq:d_aug_C_r_d_alpha_approx,eq:aug_C_r_prime_inv_approx}, we obtain
\begin{equation}
  \begin{aligned}
    [\m{J}'_F]_{1,2} & = \frac{N_{\text{OFDM}}}{2} \tr*{ \am{C}_{\ve{r}}'^{-1} \frac{\partial \am{C}_{\ve{r}}'}{\partial K_1} \am{C}_{\ve{r}}'^{-1} \frac{\partial \am{C}_{\ve{r}}'}{\partial \alpha^*} }\\
    & \approx \frac{N_{\text{OFDM}}}{2} \tr*{ \am{C}_{\ve{r}}'^{-1} \frac{\partial \am{C}_{\ve{r}}'}{\partial K_1} \am{C}_{\ve{r}}'^{-1} \begin{bmatrix} \m{0} & \m{0} \\ 2 \Re\{\m{C}_{\ve{s}}'\} & \m{0} \end{bmatrix} },
  \end{aligned}
\end{equation}
with
\begin{equation} \label{eq:J_F_arg_block_diag}
  \am{C}_{\ve{r}}'^{-1} \frac{\partial \am{C}_{\ve{r}}'}{\partial K_1} \am{C}_{\ve{r}}'^{-1} = \begin{bmatrix} \m{C}_{\ve{r}}'^{-1} \m{C}_{\ve{s}}' \m{C}_{\ve{r}}'^{-1} & \m{0} \\ \m{0} & \m{C}_{\ve{r}}'^{-*} \m{C}_{\ve{s}}'^* \m{C}_{\ve{r}}'^{-*} \end{bmatrix}.
\end{equation}
Following \cref{eq:block_matrix_mult}, the multiplication of the block-diagonal matrix $\am{C}_{\ve{r}}'^{-1} (\partial \am{C}_{\ve{r}}' / \partial K_1) \am{C}_{\ve{r}}'^{-1}$ with the off-diagonal matrix $\partial \am{C}_{\ve{r}}' / \partial \alpha^*$ inside the trace operator yields
\begin{equation}
  [\m{J}'_F]_{1,2} \approx \tr*{\begin{bmatrix} \m{0} & \m{0} \\ 2 \m{C}_{\ve{r}}'^{-*} \m{C}_{\ve{s}}'^* \m{C}_{\ve{r}}'^{-*} \Re\{\m{C}_{\ve{s}}'\} & \m{0} \end{bmatrix}} = 0,
\end{equation}
and similarly $[\m{J}'_F]_{2,1} \approx 0$.
Consequently, the \ac{FIM} becomes diagonal, and the minimum variance of $\hat{\alpha}$ is simply 
\begin{equation} \label{eq:var_alpha_hat_approx}
  \Var{\hat{\alpha}} \approx [\m{C}'_{\hat{\boldsymbol{\uptheta}}_{\alpha}}]_{2,2} \gtrsim 1 / [\m{J}'_F]_{2,2}.
\end{equation}

To obtain an analytic result for this variance, we substitute the approximated matrices \cref{eq:aug_C_r_prime_inv_approx,eq:d_aug_C_r_d_alpha_approx,eq:d_aug_C_r_d_alpha_conj_approx} in the \ac{FIM} from the optimized calculation \cref{eq:J_F_simple} leading to
\begin{equation}
  \begin{aligned}
    \label{eq:J_F_simple_22}
    [\m{J}'_F]_{2,2} & = \frac{N_{\text{OFDM}}}{2} \tr*{\am{C}_{\ve{r}}'^{-1} \frac{\partial \am{C}_{\ve{r}}'}{\partial \alpha} \am{C}_{\ve{r}}'^{-1} \frac{\partial \am{C}_{\ve{r}}'}{\partial \alpha^*} } \\
    & \approx \frac{N_{\text{OFDM}}}{2} \tr*{ \m{C}_{\ve{r}}'^{-1} (2 \Re\{\m{C}_{\ve{s}}'\}) \m{C}_{\ve{r}}'^{-*} (2 \Re\{\m{C}_{\ve{s}}'\}) }.
  \end{aligned}
\end{equation}
Leveraging the eigenvalue-based approach from the previous section, the trace in \cref{eq:J_F_simple_22} can be computed using the diagonal elements of the corresponding frequency-domain matrices. 
In particular, the eigenvalues of $\m{C}_{\ve{r}}'^{-1}$ are given by the diagonal elements $[\m{C}_{\ve{r},f}'^{-1}]_{n,n} = 1/[\m{C}_{\ve{r},f}']_{n,n} = 1/\sigma_{r,f_n}^2$, those of $\m{C}_{\ve{r}}'^{-*}$ by $[\m{C}_{\ve{r},f,\text{Img}}'^{-1}]_{n,n} = 1/[\m{C}_{\ve{r},f,\text{Img}}']_{n,n} = 1/\sigma_{r,f_n,\text{Img}}^2$, and the term $2 \Re\{\m{C}_{\ve{s}}'\} = \m{C}_{\ve{s}}' + \m{C}_{\ve{s}}'^*$ corresponds to $[\m{C}_{\ve{s},f}']_{n,n} + [\m{C}_{\ve{s},f,\text{Img}}']_{n,n} = \sigma_{s,f_n}^2 + \sigma_{s,f_n,\text{Img}}^2$.
The trace of the product of diagonal matrices is the sum of the products of their diagonal elements, resulting in
\begin{equation}
  [\m{J}'_F]_{2,2} \approx \frac{N_{\text{OFDM}}}{2} \sum_{n=0}^{\Ndft-1} \frac{(\sigma_{s,f_n}^2 + \sigma_{s,f_n,\text{Img}}^2)^2}{\sigma_{r,f_n}^2 \sigma_{r,f_n,\text{Img}}^2}.
\end{equation}
Finally, substituting this result into \cref{eq:var_alpha_hat_approx} leads to the simplified \ac{CRLB} expression:
\begin{equation} \label{eq:min_var_alpha_hat_approx_full}
  \Var{\hat{\alpha}} \gtrsim \frac{2}{N_{\text{OFDM}}} \left( 
      \sum_{n = 0}^{\Ndft - 1} \frac{
          \left(\sigma_{s,f_n}^2 + \sigma_{s,f_n,\text{Img}}^2\right)^2
        }{
          \sigma_{r,f_n}^2 \sigma_{r,f_n,\text{Img}}^2
        } 
    \right)^{-1}.
\end{equation}

Based on this equation, we would like to highlight and interpret the resulting expressions for two special cases of particular interest:
\subsubsection{Symmetric Allocation} 
A symmetric allocation (meaning the logical allocation pattern from \cref{eq:logical_alloc_pattern} fulfills $\boldsymbol{\psi} = \m{M} \boldsymbol{\psi}$) in the frequency-flat channel case $\m{H} = \m{I}$, with $\sigma_{\eta_s}^2 = 0$ and $\sigma_{\eta_r}^2 \geq 0$, yields
\begin{equation} \label{eq:min_var_alpha_hat_symm_alloc_with_noise}
  \Var{\hat{\alpha}} \gtrsim \frac{1}{2 L_s N_{\text{OFDM}}} \left(1 + \xi_r^{-1} \frac{L_s}{\Ndft}\right)^2,
\end{equation}
where $\xi_{r} = \sigma_{r}^2 / \sigma_{\eta_r}^2$ is the post-imbalance \ac{SNR}.
If we consider the zero-noise case where $\xi_r \to \infty$, while maintaining the other conditions ($\sigma_{\eta_s}^2 = 0$, $\m{H} = \m{I}$), the expression further simplifies to
\begin{equation} \label{eq:min_var_alpha_hat_symm_alloc_simplified}
  \Var{\hat{\alpha}} \gtrsim \frac{1}{2 L_s N_{\text{OFDM}}}.
\end{equation}
Note that the variance expression derived in \cite{paireder_ultra-low_2019} assumes $\boldsymbol{\psi} = \m{1}_{\Ndft}$, i.e., $L_s = \Ndft$ in the notation of this work. A direct comparison of their findings with \eqref{eq:min_var_alpha_hat_symm_alloc_simplified} reveals that their estimation variance is by a factor of two greater than the \ac{CRLB} for a small \ac{I/Q} imbalance. These findings are corroborated by the simulation results in Fig.~\ref{fig:CRLB_plot_combined}(b).

\subsubsection{Asymmetric Allocation}
An asymmetric allocation (defined by $\boldsymbol{\psi} \odot \m{M} \boldsymbol{\psi} = \ve{0}_{\Ndft}$ where $'\odot'$ denotes the Hadamard product), a frequency-flat channel, together with $\sigma_{\eta_s}^2 = 0$ (greater than zero would again resemble a non-asymmetric allocation), and $\sigma_{\eta_r}^2 \geq 0$ yields 
\begin{equation} \label{eq:min_var_alpha_hat_asymm_alloc}
  \Var{\hat{\alpha}} \gtrsim \frac{\xi_r^{-1}}{N_{\text{OFDM}} \Ndft} \left(1 + \xi_r^{-1} \frac{L_s}{\Ndft}\right).
\end{equation}
From this expression, it follows that $\xi_r \to \infty \Rightarrow \Var{\hat{\alpha}} = 0$ for an asymmetric allocation. More generally, \eqref{eq:min_var_alpha_hat_asymm_alloc} implies that $\alpha$ is directly calculable in the zero-noise case if $x[n]$ is a weighted sum of discrete-time complex exponentials $x[n]=\sum_{k\in\mathcal{S}} d[k] \exp(\j \omega_k n / N)$, provided that $k\in\mathcal{S} \Rightarrow -k\notin\mathcal{S}$ and $0 \notin\mathcal{S}$. This finding is consistent with \cite{paireder_ultra-low_2019}, which states that $\alpha$ can be directly calculated from $\ve{r}$ in the zero-noise case with a complex exponential input sequence.
When \ac{AWGN} is present, the term $\xi_r^{-1} L_s / \Ndft$ in \cref{eq:min_var_alpha_hat_asymm_alloc} implies that a single-tone complex exponential ($L_s = 1$) minimizes the \ac{CRLB} for a fixed $\xi_r$. Intuitively, concentrating the signal power on fewer subcarriers yields a higher \ac{PSD}---and thus a higher \ac{SNR}---at the information-bearing frequencies.

\section{Simulations} \label{sec:simulations}
This section compares the derived \acp{CRLB} to the performance of several blind \ac{I/Q} imbalance estimators. It shows that the small \ac{I/Q} imbalance approximation has a minimal impact on the bound, thereby establishing it as a reliable performance benchmark and enabling the evaluation of estimator optimality. To provide a concrete and representative evaluation, we adopt 5G \ac{NR} parameters, although the derived bounds apply to any \ac{CP}-\ac{OFDM} system. The simulation parameters are as follows: $1024$-\ac{QAM} modulation alphabet, $\Ndft = 4096$, $N_{\text{OFDM}} = 10$, $\sigma_{\eta_s}^2 = 10^{-2}$, $\sigma_{\eta_r}^2 = 10^{-3}$, and $\sigma_d^2 = 1$. The results are averaged over $R = 10^5$ independent simulation runs with random channel realizations according to the simplified delay profile TDLB100 \cite{3gpp.38.101-4}. In the following, $(\cdot)^{(r)}$ denotes the value of a parameter or an estimate from the $r$th simulation run. Each channel realization is normalized such that the average signal power remains unchanged by the channel ($\sigma_x^2 = \sigma_{\check{s}}^2$). Furthermore, $\alpha^{(r)}$ and $K_1^{(r)}$ are selected randomly for each simulation run such that the resulting $\text{ILR}_{\text{dB}}= \qty{-20}{dB}$. 
\begin{figure*}[!t]
  \centering
  \includegraphics[width=\linewidth]{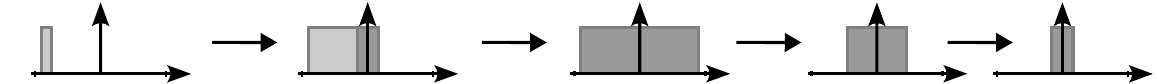}\\[4pt]
  \input{figures/CRLB_plot_combined.tex}
  \caption{\acs{I/Q} imbalance estimation CRLB ($10 \log_{10} [\m{C}_{\hat{\boldsymbol{\uptheta}}_{\alpha}}]_{2,2}$) for frequency-selective and frequency-flat channels, and \acs{MSE} performance of various estimators versus the number of allocated subcarriers. (a)~Contiguously allocated subcarriers starting from the lowest subcarrier indices. (b)~Symmetrically allocated subcarriers centered around DC. The x-axis of~(b) is reversed so that full allocation appears where both plots meet.}
  \label{fig:CRLB_plot_combined}
\end{figure*}

\subsection{Performance versus Allocation}
Fig.~\ref{fig:CRLB_plot_combined} plots three \ac{CRLB} versions: 
\textit{frequency-selective CRLB} and \textit{frequency-flat CRLB} denote the \acp{CRLB} without the small \ac{I/Q} imbalance approximation for a frequency-selective channel ($\m{H} \neq \m{I}_N$) and a frequency-flat channel ($\m{H} = \m{I}_N$), respectively. The \textit{simplified CRLB} denotes the \ac{CRLB} for a frequency-selective channel with the small \ac{I/Q} imbalance approximation. Furthermore, the performances of the \ac{RLS} estimator \cite{song_blind_2018}, the \ac{MBE} \cite{anttila_blind_2006}, and the \ac{ULC} estimator \cite{paireder_ultra-low_2019} are displayed. Note, that all estimator performances in the figures are expressed in terms of 
\begin{equation}
  \begin{aligned}
    \overline{\text{ILR}}_{\text{c,dB}} & = 10 \log_{10} \left( \frac{1}{R} \sum_{r=1}^R |\alpha^{(r)} - \hat{\alpha}^{(r)}|^2 \right) \\
    & \approx 10 \log_{10} \text{MSE}(\hat{\alpha}),
  \end{aligned}
\end{equation}
cf. \eqref{eq:ILR_c_dB}, as this is a practically relevant performance metric. The bounds in the figures are obtained by evaluating \cref{eq:Var_alpha,eq:min_var_alpha_hat_approx_full} and inserting the results into $10 \log_{10} \left( \frac{1}{R} \sum_{r=1}^R \Var{\hat{\alpha}}^{(r)} \right)$.
The pictograms above the plots illustrate the respective subcarrier allocation patterns, where darker shading indicates symmetrically allocated subcarriers and lighter shading indicates asymmetrically allocated ones.

In Fig.~\ref{fig:CRLB_plot_combined}(a), the subcarriers are allocated contiguously, beginning with a single \ac{RB} at the lowest subcarrier indices and extending to a fully allocated \ac{OFDM} symbol with $3300$ allocated subcarriers at the rightmost end of the x-axis, as illustrated by the pictograms above the plot.

In the left half of the plot, where the simulated subcarrier allocations are asymmetric, the \ac{MBE} essentially attains the three \ac{CRLB} versions. Since they practically coincide in this region, we henceforth refer to them as ``the CRLB'' whenever this is the case.
Interestingly, all three estimators exhibit a drastic performance degradation once the allocation includes symmetrically allocated subcarrier pairs (i.e., $\boldsymbol{\psi} \odot \m{M} \boldsymbol{\psi} \neq \ve{0}_{\Ndft}$). The \ac{CRLB}, although worse than for a purely asymmetric subcarrier allocation, still indicates that there is room for improvement.

A comparison of \eqref{eq:min_var_alpha_hat_symm_alloc_with_noise} and \eqref{eq:min_var_alpha_hat_asymm_alloc} confirms that asymmetric allocations are superior to non-asymmetric ones for \ac{I/Q} imbalance estimation in the low-noise case.
Motivated by this insight and the simulation results above, we implemented an ideal frequency-domain filter that removes symmetrically allocated spectral components (indicated by $\boldsymbol{\psi} \odot \m{M} \boldsymbol{\psi}$) from $\ve{r}$ prior to estimation. The resulting performances are shown as dashed lines in Fig.~\ref{fig:CRLB_plot_combined}(a). These results indicate that filtering increases the estimation performance, and in case of the \ac{MBE} practically reaches the bound for most of the plotted allocations. The remaining gap to the \ac{CRLB} compared to the asymmetric region stems from the decreased \ac{SNR} due to the removal of spectral information-bearing components. The influence of noise on estimation performance is further examined in Section~\ref{sec:perf_vs_snr}.

As mentioned, the filtered \ac{MBE} nearly attains the \ac{CRLB} for most of the subcarrier allocations. In our simulations, its deviation from the \ac{CRLB} begins to increase at $L_s \approx 2500$. As $L_s$ approaches a full allocation, the unfiltered \ac{MBE} unsurprisingly achieves better performance than the filtered version, since little asymmetric communication data remains after filtering.

Fig.~\ref{fig:CRLB_plot_combined}(b) shows the simulation results for purely symmetric allocations. As depicted by the pictograms, the rightmost data points correspond to an allocation consisting of two symmetrically allocated \acp{RB} centered around the \ac{DC} subcarrier. As $L_s$ increases, the allocation extends symmetrically around \ac{DC} towards higher frequencies, eventually reaching maximum bandwidth usage ($L_s = 3300$) at the leftmost side of the plot. At this point, the allocations of both plots coincide, as both correspond to a full allocation.

Across the plot, there is a noticeable performance gap of approximately \qtyrange{3}{20}{dB} between the frequency-selective \ac{CRLB} and the estimators, with the \ac{MBE} demonstrating the smallest gap and the \ac{RLS} showing nearly equal performance. It is important to note that the \ac{CRLB} derivation in this work assumes perfect channel knowledge, whereas the simulated estimators do not leverage this information. This likely contributes to the performance gap between the bound and the estimators. However, a gap remains even between the frequency-flat CRLB and the estimator's performances, which is attributable to noise effects, analyzed in Section~\ref{sec:perf_vs_snr}.

It is interesting to note that the frequency-flat \ac{CRLB} is consistently higher than the frequency-selective \ac{CRLB} in both Fig.~\ref{fig:CRLB_plot_combined}(a) and \ref{fig:CRLB_plot_combined}(b). In (a), this gap remains small and only reaches approximately \qty{3}{dB} at the full allocation, while for the symmetric allocations in (b), it is pronounced over almost the whole plotted subcarrier allocation range. Frequency-selective fading channels destroy the symmetry of the received signal's \ac{PSD}, thereby facilitating \ac{I/Q} imbalance estimation. 

\subsection{Performance versus Pre-Imbalance SNR} \label{sec:perf_vs_snr}
\begin{figure}[!t]
  \centering
  \input{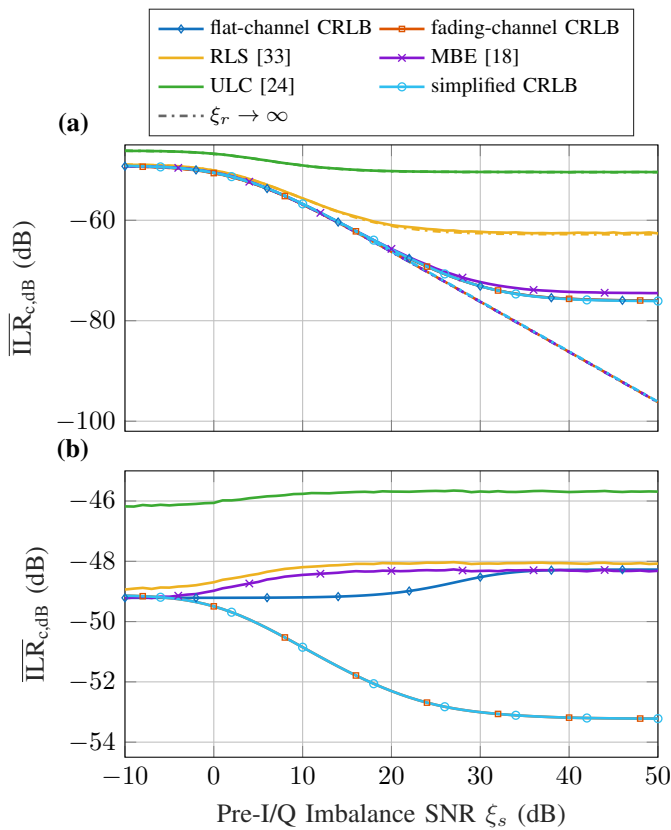}
  \caption{\acs{I/Q} imbalance estimation CRLB for frequency-selective and frequency-flat channels, and performance of various estimators versus $\xi_s$ in \unit{dB} for the maximum $L_s$ with an asymmetric allocation (top) and for a full allocation (bottom).}
  \label{fig:CRLB_SNR_plot_combined}
\end{figure}
In Fig.~\ref{fig:CRLB_SNR_plot_combined}, the three versions of the \ac{CRLB}, and the average performances of the three estimators from the literature are plotted with varying pre-\ac{I/Q} imbalance \ac{SNR} $\xi_s = \sigma_s^2 /\sigma_{\eta_s}$. For all simulation parameters other than $\sigma_{\eta_s}^2$ and the subcarrier allocation, the same values as for the results in Fig.~\ref{fig:CRLB_plot_combined} were used. Additionally, to provide further insights, we include results with zero post-\ac{I/Q}-imbalance noise ($\xi_r \rightarrow \infty$), shown as dash-dotted lines with a representative black legend entry.

The results presented in Fig.~\ref{fig:CRLB_SNR_plot_combined}(a) correspond to half of the bandwidth being allocated, which represents the maximum possible number of allocated subcarriers $L_s = 3300 / 2 - 1 = 1649$ for an asymmetric allocation. It can be seen, that the three \ac{CRLB} versions basically coincide over the whole displayed range of $\xi_s$. Interestingly, the \ac{CRLB} indicates an $\overline{\text{ILR}}_{\text{c,dB}}$ floor at high $\xi_s$ and a ceiling at low $\xi_s$. 

For the asymmetric allocation, this $\overline{\text{ILR}}_{\text{c,dB}}$ floor can be explained by the post-\ac{I/Q}-imbalance noise, whose average power was held constant across the plot. This becomes apparent from the simulation results with $\sigma^2_{\eta_r} = 0 \implies \xi_r \rightarrow \infty$ (the dash-dotted lines), where the performance floor of the \ac{CRLB} and the \ac{MBE} vanishes. Instead, the $\overline{\text{ILR}}_{\text{c,dB}}$ of the \ac{CRLB} and the \ac{MBE} decrease linearly with increasing $\xi_s$, which coincides with our simplified \ac{CRLB} for asymmetric allocations \eqref{eq:min_var_alpha_hat_asymm_alloc}, where the linear term dominates for large $\xi_{s}$. Notably, the \ac{MBE} attains the \ac{CRLB} over the whole $\xi_s$ range in this scenario.

When post-\ac{I/Q}-imbalance noise is present ($\sigma_{\eta_r} = 10^{-3}$, the solid lines), the \ac{MBE}'s $\overline{\text{ILR}}_{\text{c,dB}}$ floor is approximately \qty{1.5}{dB} higher than that of the bound. The \ac{RLS} and \ac{ULC} estimators display similar floor behavior, but their floors are approximately \qty{12}{dB} and \qty{24}{dB} higher, respectively, and therefore already occur at lower $\xi_s$ values. The \ac{RLS} and \ac{ULC} error floors remain unaffected by the removal of post-\ac{I/Q}-imbalance noise, as their solid and dash-dotted lines essentially coincide. This suggests that factors other than noise, such as approximations in their derivations, limit their performance.

At the lowest simulated $\xi_s$ value of \qty{-10}{dB}, an $\overline{\text{ILR}}_{\text{c,dB}}$ ceiling becomes apparent for the \ac{CRLB} and all simulated estimators. Similar to the symmetric allocation results in Fig.~\ref{fig:CRLB_plot_combined}(b), the \ac{ULC} estimator's ceiling is \qty{3}{dB} higher than that of the \ac{CRLB} and the other two estimators. These ceilings demonstrate that channel noise $\boldsymbol{\upeta}_s$ can also be exploited for \ac{I/Q} imbalance estimation, which constitutes a key advantage of blind estimation over pilot- or preamble-based approaches.

Fig.~\ref{fig:CRLB_SNR_plot_combined}(b) depicts the full allocation scenario. Here, the frequency-flat \ac{CRLB} is marginally lower for low $\xi_s$ than for high $\xi_s$, suggesting that \ac{WGN} is superior to \ac{QAM} symbols for estimation. Conversely, the frequency-selective \ac{CRLB} is approximately \qty{3}{dB} lower at high $\xi_s$. This occurs because the frequency-selective channel destroys the symmetry of the received signal's \ac{PSD}, thereby facilitating \ac{I/Q} imbalance estimation. Since the channel noise $\boldsymbol{\upeta}_r$ lacks this frequency selectivity, the \ac{CRLB} increases when noise dominates. At extremely low $\xi_s$ values, the \ac{MBE} meets the frequency-selective bound, while the \ac{RLS} exhibits a slightly higher $\overline{\text{ILR}}_{\text{c,dB}}$. The \ac{ULC} remains approximately \qty{1.5}{dB} above the \ac{MBE}, mirroring previous findings. Notably, a gap persists between the \ac{MBE} and the frequency-flat \ac{CRLB} in the medium $\xi_s$ range (\qtyrange{0}{30}{dB}). As in (a), the \ac{CRLB} exhibits floor and ceiling behavior at the extremes of the $\xi_s$ range.

\subsection{Performance versus ILR}
\begin{figure}[!t]
  \centering
  \input{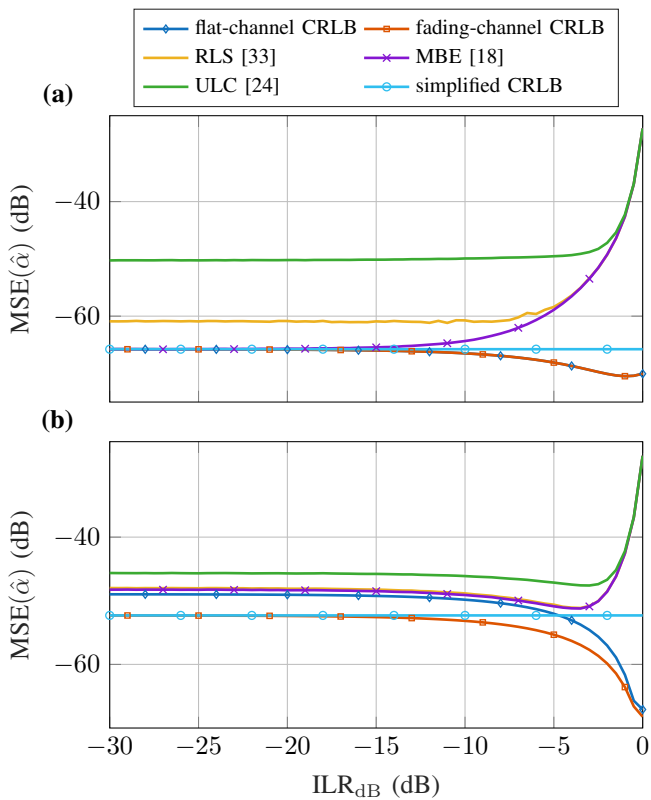}
  \caption{\acs{I/Q} imbalance estimation CRLB for frequency-selective and frequency-flat channels, and performance of various estimators versus the $\text{ILR}_{\unit{dB}}$ for a half bandwidth allocation (top) and for a full allocation (bottom).}
  \label{fig:CRLB_ILR_plot_combined}
\end{figure}
In Fig.~\ref{fig:CRLB_ILR_plot_combined}, the three versions of the \ac{CRLB} and the average performances of the three estimators are plotted versus the $\text{ILR}_{\unit{dB}}$ before compensation. Unlike the previous figures, the y-axis shows the $\text{MSE}(\hat{\alpha})$ rather than the approximate \ac{ILR} after compensation, since the small \ac{I/Q} imbalance approximation assumption for \cref{eq:min_var_alpha_hat_approx_full} is violated at high \ac{ILR} values (where $\alpha \hat{\alpha} \neq 0$). For all simulation parameters other than $\alpha$ and the subcarrier allocation, the same values as for the results in Fig.~\ref{fig:CRLB_plot_combined} were used.

As in the previous figure, the results presented in Fig.~\ref{fig:CRLB_ILR_plot_combined}(a) correspond to half of the bandwidth being allocated ($L_s = 1649$).
The flat- and frequency-selective \acp{CRLB} show that the minimum estimation accuracy slightly improves with increasing \ac{I/Q} imbalance (higher \ac{ILR}), starting from approximately \qty{-66}{dB} at low \ac{ILR} and reaching \qty{-70}{dB} at \qty{0}{dB} \ac{ILR}. This indicates that a stronger \ac{I/Q} imbalance facilitates estimation.
As expected, the simplified \ac{CRLB} is constant at approximately \qty{-66}{dB}, as it assumes $|\alpha| \to 0$. However, it accurately matches the exact \acp{CRLB} for $\text{ILR}_{\unit{dB}} \lesssim \qty{-15}{dB}$, confirming the validity of the small \ac{I/Q} imbalance approximation in the relevant range.

The estimators (MBE, RLS, ULC) exhibit the opposite trend: their performance degrades significantly at high \ac{ILR} (large imbalance). Similar to the findings in the previous subsections, the \ac{MBE} approaches the \ac{CRLB} for sufficiently low \ac{ILR} values ($\text{ILR}_{\unit{dB}} \lesssim \qty{-15}{dB}$), while \ac{RLS} and \ac{ULC} exhibit error floors.

Consistent with the previous figures, Fig.~\ref{fig:CRLB_ILR_plot_combined}(b) depicts the simulation results for the full allocation scenario.
Here, the frequency-selective \ac{CRLB} is consistently lower than the frequency-flat \ac{CRLB}, with a gap of approximately \qty{3}{dB} in the low \ac{ILR} region, reinforcing the benefit of frequency selectivity for estimation.
Similar to (a), the flat- and frequency-selective \acp{CRLB} decrease as \ac{ILR} increases, while the simplified \ac{CRLB} remains constant. The small \ac{I/Q} imbalance approximation error again becomes negligible for $\text{ILR}_{\unit{dB}} \lesssim \qty{-15}{dB}$.

The estimator performances begin to worsen at very high \ac{ILR} values ($> \qty{-3}{dB}$). At low \ac{ILR}, the \ac{MBE} and the \ac{RLS} approach the frequency-flat \ac{CRLB} but remain above the frequency-selective \ac{CRLB}.

\section{Conclusion} \label{sec:conclusion}
We presented an analytic \ac{CRLB} for blind estimation of \ac{FID} receiver \ac{I/Q} imbalance under a general \ac{OFDM} \ac{DL} signal model. The derivation rests on a \ac{CLT}-based Gaussian approximation of time-domain \ac{OFDM} samples, shown to be accurate once a moderate number of subcarriers is allocated, and on a widely linear formulation that yields a closed-form bound for the estimation of the complex-valued \ac{I/Q} imbalance parameter $\alpha$. This fills a gap left by prior bounds that rely on pilots, training, or data-dependent formulations.

Beyond the bound itself, we provided a computational optimization that reduces the cubic complexity of evaluating the inverse \ac{FIM} to linear in the \ac{DFT} size by (i) discarding \ac{CP} rows/columns with negligible information loss and (ii) exploiting eigen-decompositions of block-circulant structures for the signal covariance matrices. This makes the bound calculation practical for typical \ac{OFDM} parameterizations and enables design-space sweeps without prohibitive computational burden.

We draw the following conclusions from the simulation results. For asymmetric allocations, the \ac{MBE} essentially attains the bound, while for non-asymmetric and symmetric allocations all tested blind estimators exhibit a notable gap to the bound. A simple modification---zeroing the symmetrically allocated spectral components before estimation---closes most of this gap for non-asymmetric allocations, and drives the moment-based approach close to the CRLB over a wide range of allocations.

The extended performance analyses show that frequency selectivity in the channel improves the bound by approximately \qty{3}{dB} for full allocations compared to a flat channel. Regarding the pre-imbalance \ac{SNR}, we identified a performance ceiling at low \ac{SNR} levels, indicating that channel noise contributes useful information for blind estimation. Investigations of the \ac{I/Q} imbalance magnitude indicate that the bound decreases with increasing imbalance, and that our simplified closed-form approximation remains accurate for \acp{ILR} up to \qty{-15}{dB}.

Practically, the \ac{CRLB} provides two recommendations: (i) whenever feasible, prefer asymmetric allocations or (ii) apply pre-estimation filtering to suppress symmetrically allocated subcarrier pairs.

\bibliographystyle{IEEEtran}
\bibliography{IEEEfull,IEEEconf_local,bibliography}

\vfill

\end{document}